\documentstyle[referee,twocolumn,psfig,rotating]{mn}

\def\nn{n}

\def\lsim{~\rlap{$<$}{\lower 1.0ex\hbox{$\sim$}}}
\def\bsim{~\rlap{$>$}{\lower 1.0ex\hbox{$\sim$}}}
\def\sp{{ s}}

\def\dd{{\rm d}}
\def\ln{{\rm ln}}

\def\pmb#1{\setbox0=\hbox{#1}%
\kern-.025em\copy0\kern-\wd0
\kern.05em\copy0\kern-\wd0
\kern-.025em\raise.0433em\box0}

\begin{document}
\title[self-similar gas  collapse]
{Self-similar collapse of collisional gas in an expanding Universe}
\author[Chuzhoy \& Nusser ]{Leonid Chuzhoy
and 
Adi Nusser\\
Physics Department,
Technion, Haifa 32000, Israel\\
E-mail: cleonid@tx.technion.ac.il, adi@physics.technion.ac.il
}
\maketitle

\begin{abstract}

Similarity solutions are found for the adiabatic collapse of density
perturbations $\delta M/M \propto r^{-\sp}$ $(\sp>0)$ in a flat
universe containing collisional gas only. The solutions are obtained
for planar, cylindrical, and spherical perturbations with zero initial
pressure. For adiabatic index $\gamma\ge 4/3$, a shock develops at a
fixed fraction of the current turnaround distance.  Near the center of
a spherical perturbations with $\gamma>4/3$ and $\sp > 1/2$, the gas
is in quasi-hydrostatic equilibrium (pressure supported) and has an
asymptotic power law density profile, $\rho\sim r^{-3\sp/(\sp+1)}$,
independent of $\gamma$. For $\sp\le 1/2 $, the profile depends on
$\gamma$, the pressure is finite, the temperature decreases inward,
and gravity dominates pressure causing a continuous inward flow.
Although for $1/2<\sp<2$ the temperature decreases at the center, the
gas is pressure supported.  The pressure is finite in cylindrical
perturbations for $\sp\le 2(\gamma-1)/(3\gamma-4)$, and in planar
perturbations for any $\sp>0$.  We also derive the asymptotic
behaviour of the gas variables near the center in a universe dominated
by collisionless matter.  In such a universe, the gas in a spherical
perturbation with $\sp<2$ cannot be pressure supported and the
temperature approaches a constant near the center. The solutions and
the asymptotic behaviour are relevant for modelling the gas
distribution in galaxy clusters and pancake-like superclusters, and
determining the structure of haloes of self-interacting dark matter
with large interaction cross section.
\end{abstract}
\section {Introduction}
On scales larger than a few megaparsecs, pressure forces in the
baryonic matter in the universe are negligible, so the evolution of
dark and baryonic matter is mainly determined by gravity. On small
scales pressure becomes important and may segregate between the
evolution of baryonic and dark matter.  Pressure forces, cooling of
gas, and star formation feedback, are key ingredients in galaxy
formation.  These ingredients combine to cause differences between the
distributions of galaxies and dark matter (biasing), even on large
scales where these effects are not directly important (e.g., Kaiser
1984, Dekel \& Rees 1987, Kauffmann, Nusser \& Steinmetz 1997, Benson
et. al. 2000). On scales smaller than the Jeans length of the
photo-heated intergalactic medium (IGM), pressure forces dominate
gravity and can prevent the collapse of gas into dark haloes below a
certain mass threshold.  For haloes massive enough the temperature of
the IGM can be neglected and the gas falls into the halo. The mean
free path for collisions between gas particles inside a halo is $
\approx \left({200}/{\delta_c}\right)
\left({10^{-15}}/{\sigma}\right)(1+z)^{-3} \left( {0.1}/{\Omega_b
h^2}\right) 1.6 {\mathrm pc } , $ where $\delta_c$ is the overdensity
inside the virial radius and $\sigma$ is a typical cross section for
collisions in units of ${\mathrm cm}^{2}$. This is smaller than the
virial radius of a typical halo by a few orders of magnitude.
Therefore, on its infall into the halo, the gas is likely to form
shocks and transform its kinetic energy into heat.  The hot dense gas
can then cool to form stars which explode and inject energy into the
halo gas.  Detailed study of these processes under general conditions
is not feasible. One can aim at a global parameterization based on
general physical requirements which match observational data
(Kauffman, White, Guiderdoni 1993, Somerville \& Primack 1999, Cole
et. al. 1994).  Another route would be to study special aspects which
can be treated by either numerical or analytical methods.  Here we
focus on the collapse of the baryonic gas in an Einstein-De Sitter
universe, ignoring the gas initial temperature, cooling and heating
processes.  We assume that the collapse initiates from a symmetric
scale free density peak, and that the velocity of each shell in the
peak is taken to match the general expansion of the universe. The
energy of each shell is negative and it will expand up to a maximum
distance before it starts falling towards the center of the
perturbation.  The maximum distance is termed the turnaround
radius. Shell crossing is not allowed and the collapse can
proceed in two distinct ways, either a shock wave forms, or shells
accumulate at the center.  Which of these possibilities actually
occurs, depends on the physical conditions at the center.  If the
velocity vanishes at the center than a shock wave forms. If on the
other hand physical conditions allow a non vanishing velocity at the
center then the shells accumulate at the center (Bertschinger 1985).
Inner boundary conditions can be arranged so that a shock is
accompanied by the accumulation of central mass, but  a proper
stability analysis is needed to determine whether or not this is
possible (Bertschinger 1985). In this paper we focus on shocked
collapses without the formation of a central mass.  Since the initial
gas pressure is negligible, the collapse eventually develops in a
self-similar way where the only relevant scale at any time is the
radius of the shell at maximum expansion.  Bertschinger (1985), and
Forcada-Miro \& White (1997) have studied similarity solutions in
spherically symmetric perturbations with initial relative mass excess
$\delta M/M\propto r^{-3}$, and $r^{-2}$, respectively.  Here we
derive similarity solutions in planar, cylindrical, and spherical
geometries, for the collapse of a perturbation with $\delta M/M\propto
r^{-\sp}$ for any $\sp >0$ and adiabatic index $\gamma\ge 4/3$.

In section \ref{equations} we write the equations of motion for
symmetric perturbations in planar, cylindrical, and spherical
geometry.  In section \ref{asymptotic} we discuss the asymptotic
behaviour of the fluid variables near the center, in the case of
shocked collapse.  In section \ref{numerical} we present results of
numerical integrations of the equations. In section 
\ref{asymptotic:dm} we derive asymptotic behaviour of the fluid variables in a
universe dominated by collisionless dark matter.  In section
\ref{discussion} we conclude with a discussion of the results and
their potential astrophysical consequences.

\section{The equations}
\label{equations}

We write the Newtonian equations of motion governing the adiabatic
collapse of symmetric perturbations in a collisional fluid (gas) of
adiabatic index $\gamma$ and zero initial pressure. Except section
\ref{asymptotic:dm}, we restrict the analysis here and throughout to
the collapse in a flat universe containing collisional gas only.  The
initial gas pressure is zero, so the expansion scale factor of the
universe is $a(t)\propto t^{2/3}$, the Hubble function is
$H(t)=2/(3t)$, and the background density is $\rho_c=3H^2/(8\pi
G)=1/(6\pi G t^2)$.
 
Denote by $r$ and $\upsilon\equiv \dd r/ \dd t$ the physical position
and velocity of a gas shell, where $ r=0$ is the symmetry center of
the perturbation.  Further, let $\rho(r,t)$ and $p(r,t)$ be the gas
density and pressure at $r$. As in Fillmore \& Goldreich (1984) 
define the mass within a distance $r$ from the
symmetry center  by $m(r,t)=\int_0^r x^{\nn-1}\rho(x,t)\dd
x$, where $\nn=1,2$, and 3 refer, respectively, to planar,
cylindrical, and spherical perturbations.
The mass within a fixed shell varies with time like $m \sim
t^{-2(3-\nn)/3}$, because of the Hubble expansion along $3-\nn$ of the
axes.  In this notation, the equations of motion are, the continuity
equation, \def\ff{\frac{2(3-\nn)}{3}}
\begin{equation}
\frac{\dd(\rho t^{\ff})}{\dd t}=-t^\ff \rho r^{1-\nn}\partial_r(r^{\nn-1}\upsilon) \; ,
\label{eom1}
\end{equation}
Euler,
\begin{equation}
\frac{\dd \upsilon}{\dd t}-\frac{2}{9}\frac{3-\nn}{\nn}\frac{r}{t^2}
=-\frac{\partial_r p}{\rho}- \frac{4\pi G m}{r^{n-1}} 
\; ,
\label{eom2}
\end{equation}
adiabatic condition,
\begin{equation}
\frac{\dd}{\dd t}(p\rho^{-\gamma})=0 \; ,
\label{eom3}
\end{equation}
and the relation,
\begin{equation}
\partial_r m= r^{\nn -1}\rho \; .
\label{eom4}
\end{equation}

These equations are satisfied everywhere except at the shock where the
fluid variables are described by jump conditions obtained from mass,
momentum, and energy conservation.  The initial conditions leading to
self-similar collapse are specified at an early time close to zero,
$t_i$, as
\begin{eqnarray}
\label{inid} 
\frac{\delta M}{M}&=&\left(\frac{r}{r_0}\right)^{-\sp} \; , \\
\label{inivel} \upsilon(r,t_i)&=&\frac{2}{3t_i}r \; , \\
\label{inip} p(r,t_i)&=&0 \; ,
\label{inic}
\end{eqnarray}
where $\delta M/M$ is the mean density contrast interior to $r$, and
$\sp>0$.  For cosmological initial conditions the initial density
contrast must be tiny, so we will be interested in the solution in the
region $r\gg r_0$.  A perturbation with $\sp > \nn$ can be realized by
placing a high narrow positive density peak at the center ($\ll r_0$)
of a symmetric void with local density contrast $\sim (-r^{-\sp})$.
The condition (\ref{inivel}) means that a gas shell at $r$ moves
initially with the general universal expansion.  This condition can be
relaxed to allow for a non vanishing initial zero peculiar velocity
according to late time linear theory (e.g., Peebles 1980).  However,
this does not affect the details of the collapse (Peebles 1980,
Bertschinger 1985), so we use (\ref{inivel}) which is commonly adopted
in the literature.  Bertschinger (1985) and White \& Forcada (1997),
respectively, considered the collapse of spherical perturbations with
$\sp=3$, and $\sp=2$.

The equations of motion (\ref{eom1}--\ref{eom4}) together with the
initial conditions (\ref{inid}--\ref{inip}) are insufficient to
completely determine the evolution of the perturbation. Still missing
is an inner boundary condition specifying the velocity and mass at
$r=0$, for $t\ge t_i$.  For a shock to develop without the
accumulation of a central mass (a black hole for $n=3$) we must have
$v(r=0,t\ge t_i)=0$ and $m(r=0,t\ge t_i)=0$.  Relaxing the condition
$V(0)=0$ leads to a non-vanishing central mass with or without the
presence of a shock.

In a critical density universe ($\Omega=1$) the only length scale
relevant to the collapse is the scale of non-linearity.  At any time,
$t$, this scale can be defined as the distance of the shell at the
maximum expansion, i.e., the shell with $\upsilon=0$ (Gunn 1977,
Fillmore \& Goldreich 1984, Bertschinger 1985). This radius is termed
the current turnaround radius, $r_{ta}(t)$. Starting from tiny initial
density contrast, the mean overdensity (density in units of $\rho_c$)
interior to $r_{ta}(t)$ is a fixed number independent of time.  For
time $t\gg t_i$, when shells with $r\gg r_0$ reach their turnaround,
the collapse develops a self-similar behaviour that depends on $r$ and
$t$ through the combination $\lambda=r/r_{ta}$.  The turnaround radius
$r_{ta}(t)$ is given by (e.g., Fillmore \& Goldreich 1984),
\begin{equation}
r_{ta}=r_0\frac{C_x}{C_t^{3\alpha/2}} \left(\frac{t}{t_i}\right)^{\alpha} \quad ; \quad \alpha=\frac{2}{3}\frac{\sp+1}{\sp}
\end{equation}
where,
\begin{equation}
C_x=\frac{5}{12}, 0.741, 1 \quad ;\quad
C_t=\frac{5}{6}, 1.386, \left(\frac{3\pi}{4}\right)^{2/3} \; ,
\end{equation}
for $\nn=1,2,$ and 3, respectively. The turnaround radius grows faster
than the scale factor $a\sim t^{2/3}$. This is because the mass, $\sim
\rho_c r_{ta}(t)^3$, interior to $r_{ta}$ must grow with time while
the mass, $\sim \rho_c a^3(t)$, inside a fixed shell in a homogeneous
universe is constant. For $\sp <2$ the turnaround radius grows faster
than $t$ reaching the horizon scale in finite time. When this happens
relativistic description must be used and $r_{ta}$ ceases to be the
only scale in the problem (Fillmore \& Goldreich 1984).

The equations can be cast into a non-dimensional
form using the scaled variables $V(\lambda)$, $D(\lambda)$,
$P(\lambda)$, and $M(\lambda)$ defined by (Bertschinger 1985),
\begin{eqnarray}
\label{scalev}\upsilon(r,t)&=&\frac{r_{ta}}{t}V(\lambda)\\
\label{scaled}\rho(r,t)&=&\rho_c D(\lambda)\\
\label{scalep}p(r,t)&=&\rho_c\left(\frac{r_{ta}}{t}\right)^2 P(\lambda)\\
\label{scalem}m(r,t)&=&\frac{1}{3}\rho_c r_{ta}^\nn M(\lambda) \; .
\end{eqnarray}
 Expressed in terms of these variables, the equations
(\ref{eom1}-\ref{eom4}) become, respectively,
\begin{equation}
\label{a1}
\left(V-\alpha\lambda\right)D'+\left(\frac{\nn-1}{\lambda}V+V'-
\frac {2\nn}{3}\right)D=0 \; , 
\end{equation}
\begin{equation}
\label{a2}\left(\alpha-1\right)V+\left(V-\alpha\lambda\right)V'
-\frac{2}{9}\frac{3-\nn}{\nn}\lambda
=
-\frac{P'}{D}-\frac{2}{9}\frac{M}{\lambda^{\nn-1}} \; , 
\end{equation}
\begin{equation}
\label{a3}\left(\gamma\frac{D'}{D}-
\frac{P'}{P}\right)\left(V-\alpha\lambda\right)=2\left(\alpha-2+\gamma\right)
\; ,
\end{equation}
\begin{equation}
\label{a4}M'=3\lambda^{\nn-1}D \; ,
\end{equation}
where the prime symbol denotes derivatives with respect to $\lambda$.

We will mainly be concerned with solutions for shocked collapse with
vanishing mass at the center.  The inner boundary condition
appropriate for this collapse are vanishing mass and velocity at
$\lambda=0$, i.e.,
\begin{equation}
V(0)=0 \quad {\rm and} \quad M(0)=0\; ,
\label{innerb}
\end{equation} 
Self-similarity implies that the shock appears at fixed
$\lambda=\lambda_s=r_s/r_{ta}$, so the physical radius of the shock
$r_s\propto t^\alpha$ and its non-dimensional speed is $(r_{ta}/t)^{-1}(\dd
r_s/\dd t)= \alpha\lambda_s$.  At the surface of the shock the fluid
variables satisfy the jump conditions obtained from mass, momentum,
and energy conservation. In terms of the non-dimensional fluid
variables, the jump conditions appropriate for an adiabatic shock are,
\begin{eqnarray}
\label{jump1}
V^+&=&\alpha\lambda_s+\frac{\gamma-1}{\gamma+1}(V^- -\alpha\lambda_s)\; ,\\
\label{jump2}D^+&=&\frac{\gamma+1}{\gamma-1}D^- \; ,\\
\label{jump3}P^+&=&\frac{2}{\gamma+1}D^-(V^--\alpha\lambda_s)^2\; ,\\
\label{jump4}M^+&=&M^- \; ,
\end{eqnarray}
where the superscripts of the minus and plus signs refer to pre- and
post-shock quantities. In employing energy conservation we have taken
$\frac{P}{D(\gamma-1)}$ as the non-dimensional internal energy per
unit mass.

In section \ref{numerical} we will find numerical solutions satisfying
the requirements for shocked collapse without a central mass.  Except
spherical perturbations with $\gamma=4/3$ only one value
$\lambda_s$ can yield solutions satisfying these requirements.
Spherical perturbations with $\gamma=4/3$ allow a range of values for
$\lambda_s$. Before presenting the numerical solutions we derive in
the next section the asymptotic behaviour of the fluid variables near
the center, and two integrals of motion which will be used as a check
on the numerical solutions.
\section{Integrals of motion and asymptotic behaviour near the center} 
\label{asymptotic}
Solutions to (\ref{a1}--\ref{a4}) with the appropriate jump and
boundary conditions for all $\lambda $ will be found by numerical
integration.  We present here an analytic treatment of the equations
to derive the asymptotic behaviour of the fluid variables near $r=0$,
and two integrals of motion (e.g., Bertschinger 1983, 1985).  
We restrict the analysis shocked collapses satisfying the inner
boundary condition (\ref{innerb}).
 All fluid variables can be expressed in terms of an auxiliary
function $K(\lambda) \equiv \exp\int^\lambda_0\frac{\dd x}{V(x)-\alpha
x} $ as follows.
\begin{eqnarray}
\label{b6}V(\lambda)&=&\frac{K}{K'}+\alpha\lambda \; ,\\
\label{b3}\frac{D(\lambda)}{D_0}&=&
-\lambda^{1-\nn}{K'}K^{-1+\nn(2/3-\alpha)} \; ,\\
\label{b0}\frac{P(\lambda)}{P_0}&=& 
-\lambda^{(1-\nn)\gamma}
K^{4+(2\nn/3-3)\gamma -\alpha(2+\nn \gamma)} K'^\gamma \; ,\\
\label{b5}\frac{M(\lambda)}{D_0}&=& 
\frac{9 K^{\nn(2/3 -\alpha)}  }{\nn(3\alpha-2)} \; ,
\end{eqnarray}
where $D_0$ and $P_0$ are constants.  The fluid variables in
(\ref{b6}--\ref{b5}) satisfy the non-dimensional equations (\ref{a1}),
(\ref{a3}), and (\ref{a4}) for any functional form of $K$.  The
function $K$ is then specified by only one equation, the
non-dimensional Euler equation (\ref{a2}).

Two integrals of motion can immediately be found from
(\ref{b6}--\ref{b5}). These are the mass and entropy integrals of
motion (Bertschinger 1983, 1985),
\begin{eqnarray}
\label{int:mass} 
M&=&\frac{9}{\nn(2-3\alpha)}D(V-\alpha\lambda)\lambda^{\nn-1} \; ,\\
\label{int:ent} 
P D^{-\gamma}M^\zeta&=&{\rm const} \quad ; \quad  
\zeta=\frac{6}{\nn}\frac{\alpha+\gamma-2}{2-3\alpha} \; ,
\end{eqnarray} 
where all fluid variable are evaluated at any $\lambda$ inside the
shock.  Since $\zeta <0$, the entropy integral of motion means that
the entropy $\propto \ln(PD^{-\gamma})$ is an increasing function of
the mass and hence $\lambda$.

The auxiliary function greatly simplifies the derivation of the
asymptotic behaviour of the non-dimensional fluid variables near
$\lambda=0$.  Since $V(0)=0$ and $M(0)=0$, equations (\ref{b6}) and
(\ref{b5}) imply that, to first order, $K(\lambda)$ must approach
\begin{equation}
K(\lambda)= \lambda^{1/(V_0-\alpha)} 
\end{equation}
as $\lambda\rightarrow 0$, where $V_0$ is an arbitrary constant.
Substituting this expression for $K$ in (\ref{b6} -- \ref{b5}), yields
\begin{eqnarray}
\label{c4}
V&=&V_0\lambda \; , \\ 
M&=&\frac{9D_0}{n(3\alpha-2)} \lambda^{\delta+\nn} \; ,\\ 
D&=&\frac{D_0}{\alpha-V_0}\lambda^\delta\; ,\\ 
P&=&\frac{P_0}{(\alpha-V_0)^\gamma}\lambda^\eta\; ,
\end{eqnarray}
and the asymptotic exponents are expressed in terms of the coefficient
$V_0$ as
\begin{eqnarray}
\label{expdelta}\delta&=&\frac{\nn(2-3V_0)}{3(V_0-\alpha)}\; , \\
\label{expeta}\eta&=&
\frac{4-2\alpha-2\gamma+(\frac{2}{3}-V_0)\nn\gamma}{V_0-\alpha} \; .
\end{eqnarray} 
These relations have been obtained without using the Euler equation
(\ref{a2}). 
In order to determine the exponents uniquely we use 
the  Euler equation which  adds the following constraints,
\begin{eqnarray}
\eta=2(\delta+1),\; {\rm for}\;  \eta<0 \; , \nonumber \\
\label{expeuler}2(\delta+1)\geq0, \; {\rm for}\;  \eta=0 \; .
\end{eqnarray} 
A value $\eta >0$ gives $P_0<0$ and so, by (\ref{int:ent}), a negative
entropy integral of motion.  But the jump condition (\ref{jump3})
gives positive pressure, $P^+$, just behind the shock and since the density
and mass are also positive, the entropy integral (\ref{int:ent}) must
be positive. So we rule out $\eta>0$.
 
If the solution to (\ref{expdelta}), (\ref{expeta}), and 
(\ref{expeuler})  is $\eta <0$ then the Euler equation 
also provides the following  constraint on the coefficients $P_0$ and $D_0$,
\begin{equation}
\label{eqp0} P_0=
\frac{2D_0^2\alpha(V_0-\alpha)^{\gamma-1}}{\nn\eta(3\alpha-2)} \; .
\end{equation}
Table 1 lists the values of $V_0$, $\eta$, and $\delta$ in all
cases. The dashed curves in figure \ref{fig:delta} are a graphical
representation of the density exponent $\delta$ versus $\sp$ for
$\gamma=5/3$.  In planar geometry, $\nn=1$, the only possible solution
to equations is $\eta=0$ and the pressure is finite everywhere.  In
cylindrical geometry, $n=2$, we have $\eta<0$ only for
$\alpha<\frac{5\gamma-6}{3(\gamma-1)}$, so it must be zero for other
values of $\alpha$.  In spherical geometry, $n=3$, if $\alpha<2$ and
$\gamma>4/3$ then $\eta=2(1-\frac{2}{\alpha})$, and $V_0=0$ meaning
that $V(\lambda)\propto\lambda^\nu$, where $\nu>1$.  A second order
expansion gives
$\nu=1+{2\alpha}^{-1}\left[(4-5\alpha)\pm\gamma^{-1/2}\sqrt{{8\alpha(\gamma-8)+
16\gamma+\alpha^2(32+\gamma)}}\right]$, where only one of the roots is
$\nu>1$.  In the limit of either $\alpha\rightarrow 2$ or
$\gamma\rightarrow 4/3$, we have $\nu\rightarrow 1$.

For $n=3$ and $\gamma=4/3$ the relations (\ref{expdelta}),
(\ref{expeta}), and (\ref{expeuler}) allow multiple solutions for
$V_0$ and consequently for $\lambda_s$.  A second order calculation
gives the upper limit $V_0< 4-5\alpha$.  Solutions exist for any
positive $\lambda_s$ smaller than a maximal value which corresponds to
the upper limit on $V_0$. This means that for $\gamma=4/3$ a shocked
collapse cannot be accompanied by the presence of a non-vanishing mass
at the center. In the table we list the asymptotic constants
corresponding to $V_0=4-5\alpha$, i.e., the maximal value of
$\lambda_s$ for which a shocked collapse occurs.  Bertschinger (1984)
does not mention that there are solutions for shocked without a
central mass for a range of $\lambda_s$. His numerical solution with
$\gamma=4/3$ seems to correspond to the maximal $\lambda_s$ and yields
$\eta=-3.2$ and $\delta=-2.6$, instead of $-3$ and $-2.5$ as listed in
table 1.

We now examine how the dimensional density, $\rho(r,t)=\rho_c D$,
varies with time near the center. Using $\rho_c\sim t^{-2}$, the first
order expression $D\sim (r/r_{ta})^\delta$, and $r_{ta}\sim t^\alpha$,
we find $\rho(r,t)\sim r^\delta/t^{2+\alpha\delta}$.  When
$2+\delta\alpha=0$ the density is constant with time.  In spherical
geometry a time independent density is equivalent to $V_0=0$.  Because
of the expansion in the $\nn-1$ directions, a vanishing $V_0$ in
planar and cylindrical geometries leads to $\rho\sim
t^{-2-\alpha\delta} \sim t^{2(\nn-1)/3}$. In all geometries a
vanishing $V_0 $ indicates that, to first order in the asymptotic
expansion, the gas is pressure supported in hydrostatic equilibrium.
According to the table, spherical perturbations have $V_0=0$ only for
$s>1/2$.  Note that for $2>\sp>1/2$ the asymptotic density is constant
even though the temperature $\sim P/D \sim r^{\eta-\delta}$ decreases
inward.  For $\sp<1/2$, the density increases with time.  According to
table 1, planar and cylindrical perturbations have $V_0=0$ only at
$\sp=1/(2-\gamma)$ and $\sp=2/(4-\gamma)$, respectively.
 
\section{Numerical integration}
\label{numerical}
We present results of the numerical integration of the non-dimensional
equations of motion. The numerical solutions shown here describe
shocked collapses without a central mass.  Outside the shock the fluid
variables are given from the solution for collapse with zero pressure
(Zel'dovich 1970, Peebles 1980, Fillmore \& Goldreich 1984).

The shock position $\lambda_s$ is unknown a priori.  We have to
find its value such that the fluid variables satisfy the equations of
motion, (\ref{a1}--\ref{a4}), the jump conditions,
(\ref{jump1}--\ref{jump4}), and the inner boundary condition,
(\ref{innerb}).  Assuming that the pre-shock variables are given from
the zero pressure solution, the value of $\lambda_s$ can be found as
follows (Bertschinger 1985, Forcada--Miro \& White 1997).  For an
assumed value for $\lambda_s$, we obtain the post-shock variables
using the jump conditions. We then integrate the equations of motion
from $\lambda_s$ inward to $\lambda=0$, and tune $\lambda_s$ so that
the solution gives $V=0$ and $M=0$ at $\lambda=0$.  In all numerical
solutions we find that if $M(0)=0$ for a given
$\lambda_s$ then $V(0)=0$, and vice verse.

The zero pressure solutions in planar and spherical geometries are
known analytically ( Zel'dovich 1970, Fillmore \& Goldreich 1984) so
the pre-shock fluid variable can be found directly, for an assumed
$\lambda_s$. However in cylindrical geometry an analytic solution is
not available and we numerically integrate the equations with zero
pressure to obtain the pre-shock quantities.  In practice we use
numerical integration also in planar and spherical geometries.  It is
convenient to integrate the zero pressure equations from $\lambda=1$,
i.e., the turnaround radius, to $\lambda_s$. At $\lambda=1$ the fluid
variables are
\begin{equation}
 V(1)=0 \quad ,\quad P(1)=0 \; ,
\end{equation}
\begin{equation}
 D(1)=\frac{1}{\sp+1}\left(\frac{C_t}{C_x}\right)^\nn
\quad ,\quad M(1)=\frac{3}{\nn}\left(\frac{C_t}{C_x}\right)^\nn
\; ,
\end{equation} 
where the expressions for $D(1)$ and $M(1)$ were obtained from the zero
pressure solution (e.g., Fillmore \& Goldreich 1984).

The sequence of figures (\ref{fig:dens}--\ref{fig:term}) shows the
numerical solutions for the over density $D$, the pressure $P$, the
velocity $V$, and the thermal energy $U=\frac{P}{D(\gamma-1)}$ in
spherical, cylindrical and planar geometries for several values of
$\sp$, all as a function of $\lambda$. All curves are obtained from
the solutions with $\gamma=5/3$.  The solid curve in each plot
corresponds to $\sp=\nn$.  The sudden change in the fluid variables
indicate the location of the shock.  All numerical solutions satisfy
the integrals of motion (\ref{int:mass}) and (\ref{int:ent}) up to the
numerical accuracy and agree with asymptotic behaviour of the previous
section. The logarithmic slopes of all curves match the corresponding
values listed in table 1. Spherical perturbations with $\sp=3$ and $
\sp=2$ were, respectively, analyzed by Bertschinger (1984) and
Forcada--Miro \& White (1997).  The agreement between their solutions
with $\gamma=5/3$ and ours is excellent.  In all geometries the density
and pressure in the shock are higher for larger $\sp$.  Figures
\ref{fig:vel} of the velocity and figure \ref{fig:term} of the internal
energy $U=\frac{P}{D(\gamma-1)}$ demonstrate that particles are
decelerated at the shock converting most of their kinetic energy into
heat.  The velocity near the center of cylindrical perturbations can
be positive inside the shock, hence the linear vertical scale in
velocity plot in this case.

In figure \ref{fig:lam} we plot the location of the shock
$\lambda_s$ as a function of $\sp$ for several values of $\gamma$. The
shock location always increases with $\sp $ and agrees with the values
obtained by Bertschinger (1985) and Forcada-Miro \& White (1997) for
spherical perturbations with $\sp=3$ and $\sp=2$, respectively.  A
special case is   spherical collapse with $\gamma=4/3$.  Here a
solution for shocked collapse without a central point mass is possible
for any $\lambda_s $ less than a maximal value $ \lambda_c$. In this
case we plot the maximal value $\lambda_c$.  For $\sp \gg 1 $,
$\lambda_s$ can exceed unity  increasing to  a finite value as $\sp
\rightarrow \infty$.

The variation of the fluid variables attached to a given fluid
elements are also of interest. Using the velocity obtained from the
self-similar solutions we can calculate the trajectory $r(t)$ of a
fluid element (particle) as a function of time. The fluid variable
associated with the particle at any time can then be obtained by
interpolating the solutions at the particle's position at that time.
The three panels from top to bottom in figure \ref{fig:traj} show,
respectively, the trajectory, density, and pressure of a particle as
obtained from the solutions with $\gamma=5/3$ for three values of
$\sp$. The time axis in all panels is $t/t_{ta}$ where $t_{ta}$ is the
time at which the particle reached its maximum expansion (the
turnaround time). The particle position, $r$, density $\rho$, have
been scaled by their values at $t_{ta}$, $r'_{ta}=r(t_{ta})$, and
$\rho=\rho(t_{ta})$, respectively.  The pressure, $p$, has been scaled
by its value at the shock $p_s$.  With this scaling the curves in the
figures become valid for all the particles.
The particle trajectory for $s=3$ (solid line, top panel) agrees with
the corresponding curve in figure 4 of Bertschinger (1985).  As
expected from the asymptotic solution the particles for $\sp=3$ and
$\sp=10$ tend to settle at a physical distance which is a fixed
fraction of their turnaround radii.  The decay of the trajectory for
$\sp=1/4$ at late time can be evaluated using the asymptotic
expansion. Near the origin $V(\lambda)=V_0\lambda$, so $ \dd r/ \dd t
=(r_{ta}/t)V_0 \lambda=- r/t $ and $r\sim t^{V_0}$.  According to
table 1 $V_0=-8/15$ for $\sp=1/4$ and $\gamma=5/3$ so $r$ decays like
$t^{-8/15}$.
The density (middle panel) and pressure (bottom) curves for $\sp=3$
and $\sp=10$ flatten at late times, consistent with the settling of
the particles to a constant $r$.

\section{Asymptotic behaviour in a Universe dominated by collisionless matter}
\label{asymptotic:dm}

\begin{figure}
\centering
\mbox{\psfig{figure=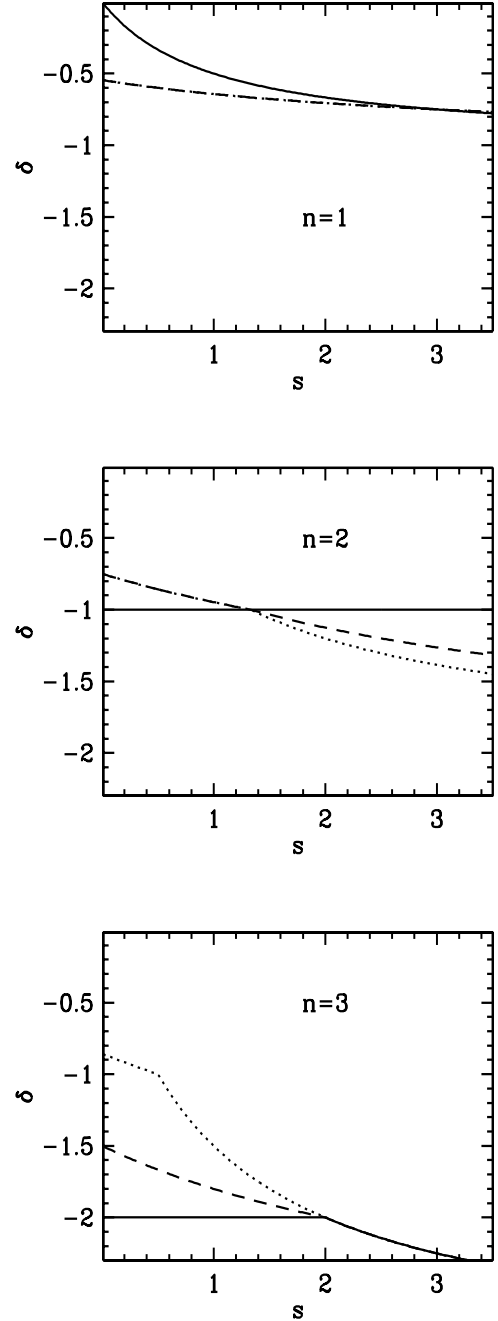,height=8.0in,width=5.in}}
\caption{The density asymptotic exponent versus $\sp$ for $\nn=1$, 2,
and 3. The dashed and dotted curves are the gas exponent for collapse
with and without collisionless matter, respectively.  The curves are
computed according to tables 1 and 2 with $\gamma=5/3$. For $\nn=2$
the two curves overlap.  The solid curve is the asymptotic exponent,
$\tilde \delta$, of the collisionless matter given by equations
(\ref{fg1}--\ref{fg3}) (Fillmore \& Goldreich 1984). }
\label{fig:delta}
\end{figure}

\begin{figure}
\centering
\begin{sideways}
\mbox{\psfig{figure=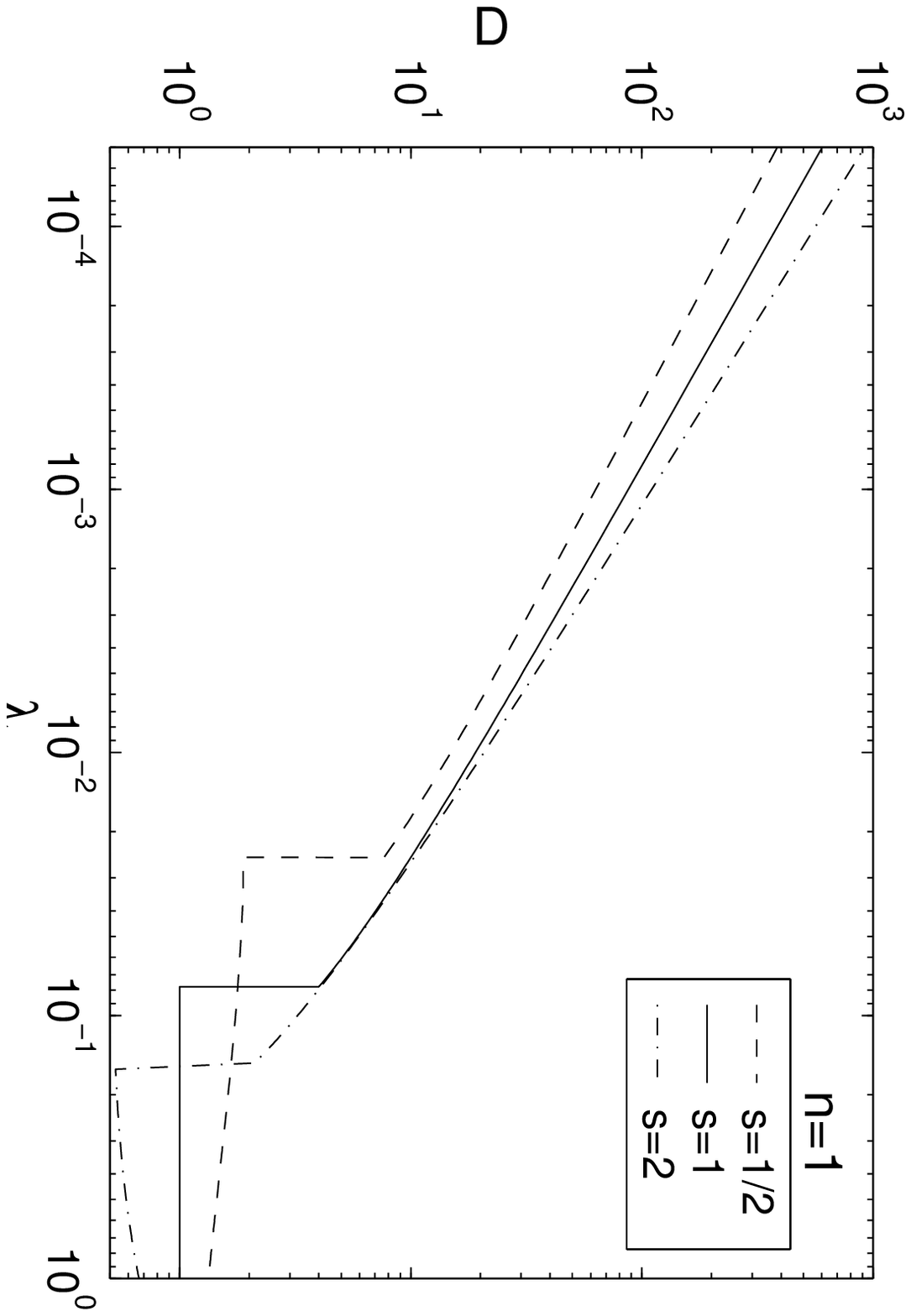,height=3.0in,width=2.5in}}
\mbox{\psfig{figure=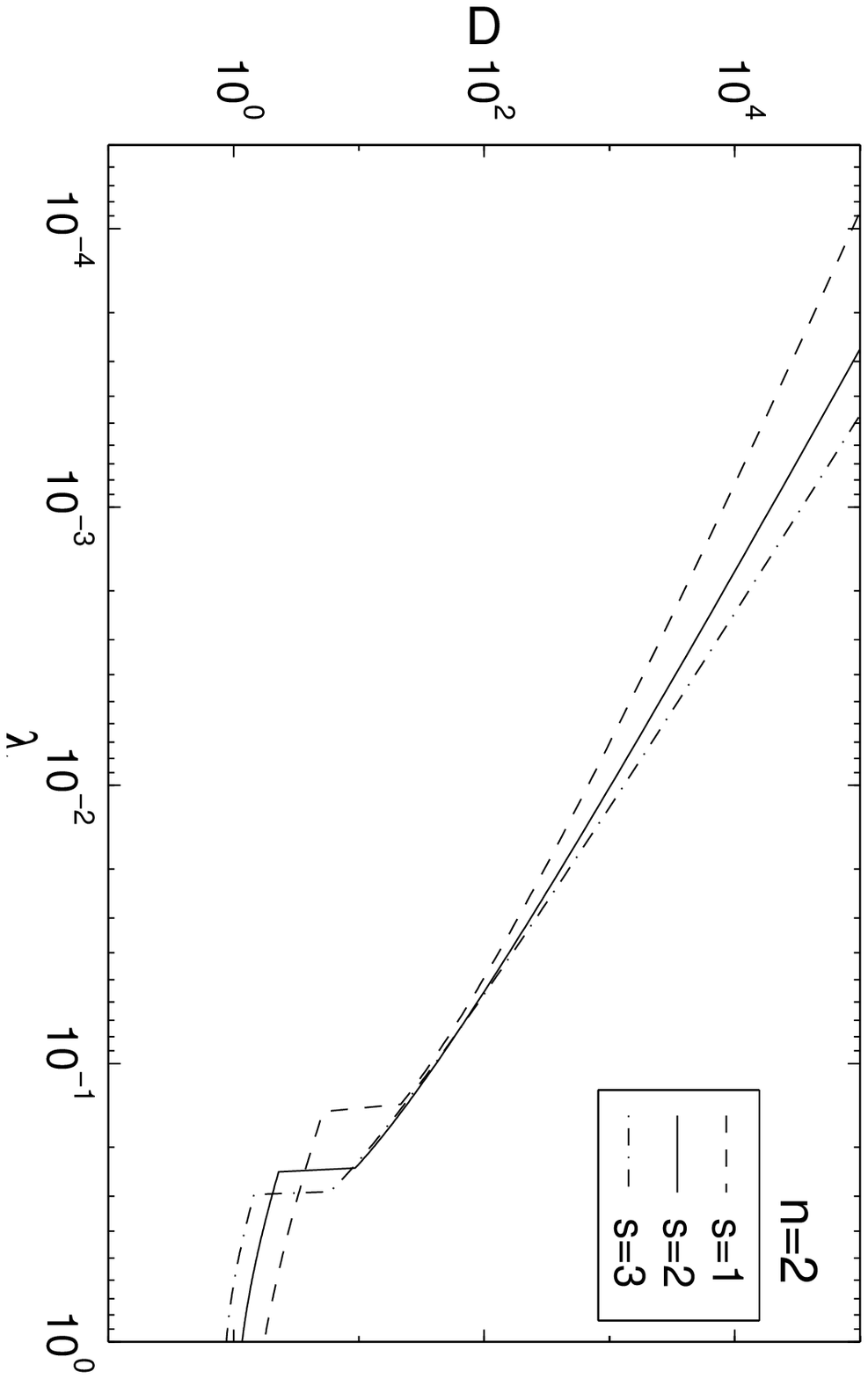,height=3.0in,width=2.5in}}
\mbox{\psfig{figure=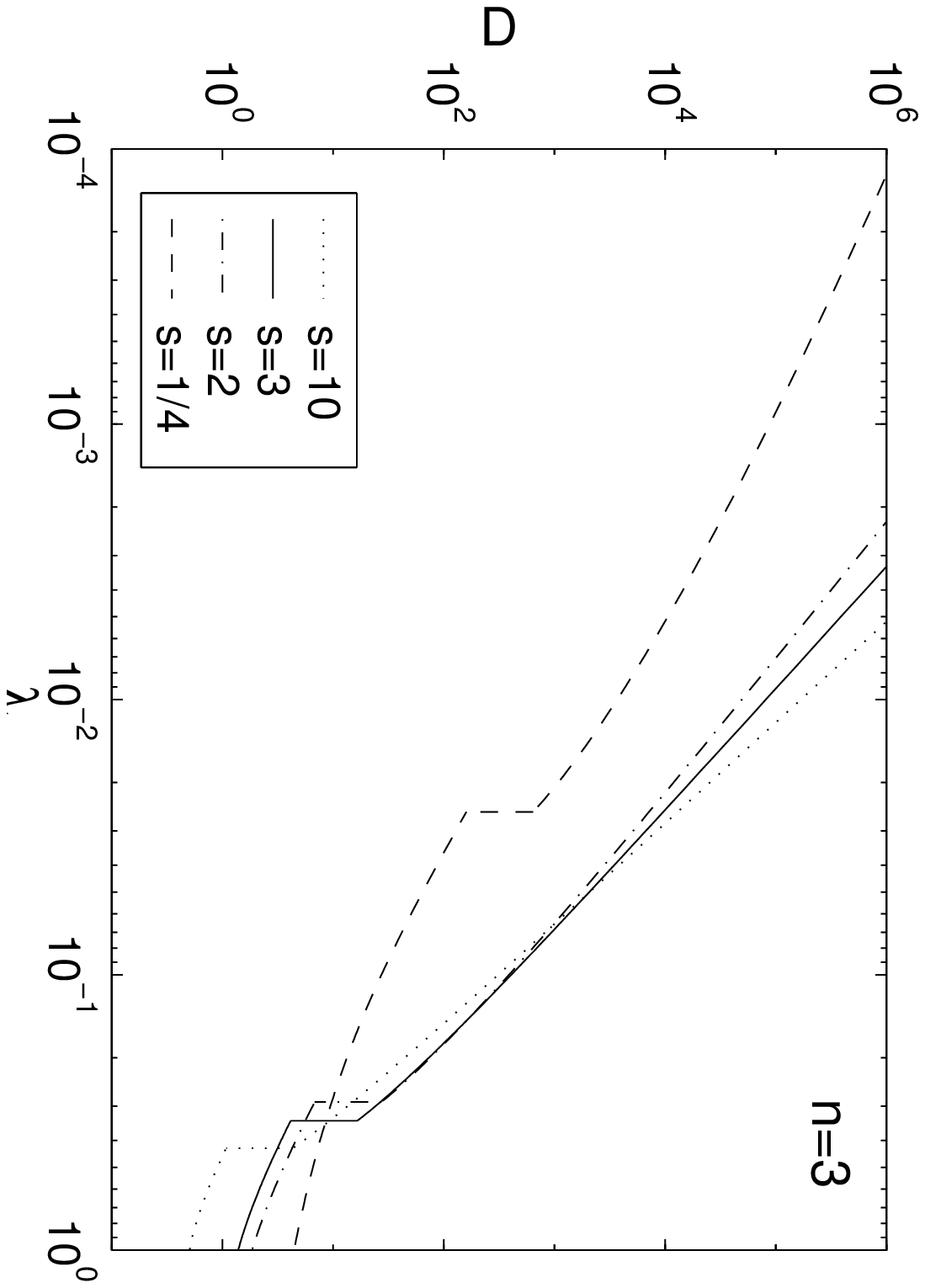,height=3.0in,width=2.5in}}
\end{sideways}
\caption{The over density $D=\rho/\rho_c$ as a function of $\lambda$
for various values of $\sp$.  
This and  figures \ref{fig:pres}--\ref{fig:term}
show fluid variables  obtained from  numerical
solutions with $\gamma=\frac{5}{3}$.  }
\label{fig:dens}
\end{figure}

\begin{figure}
\centering
\begin{sideways}
\mbox{\psfig{figure=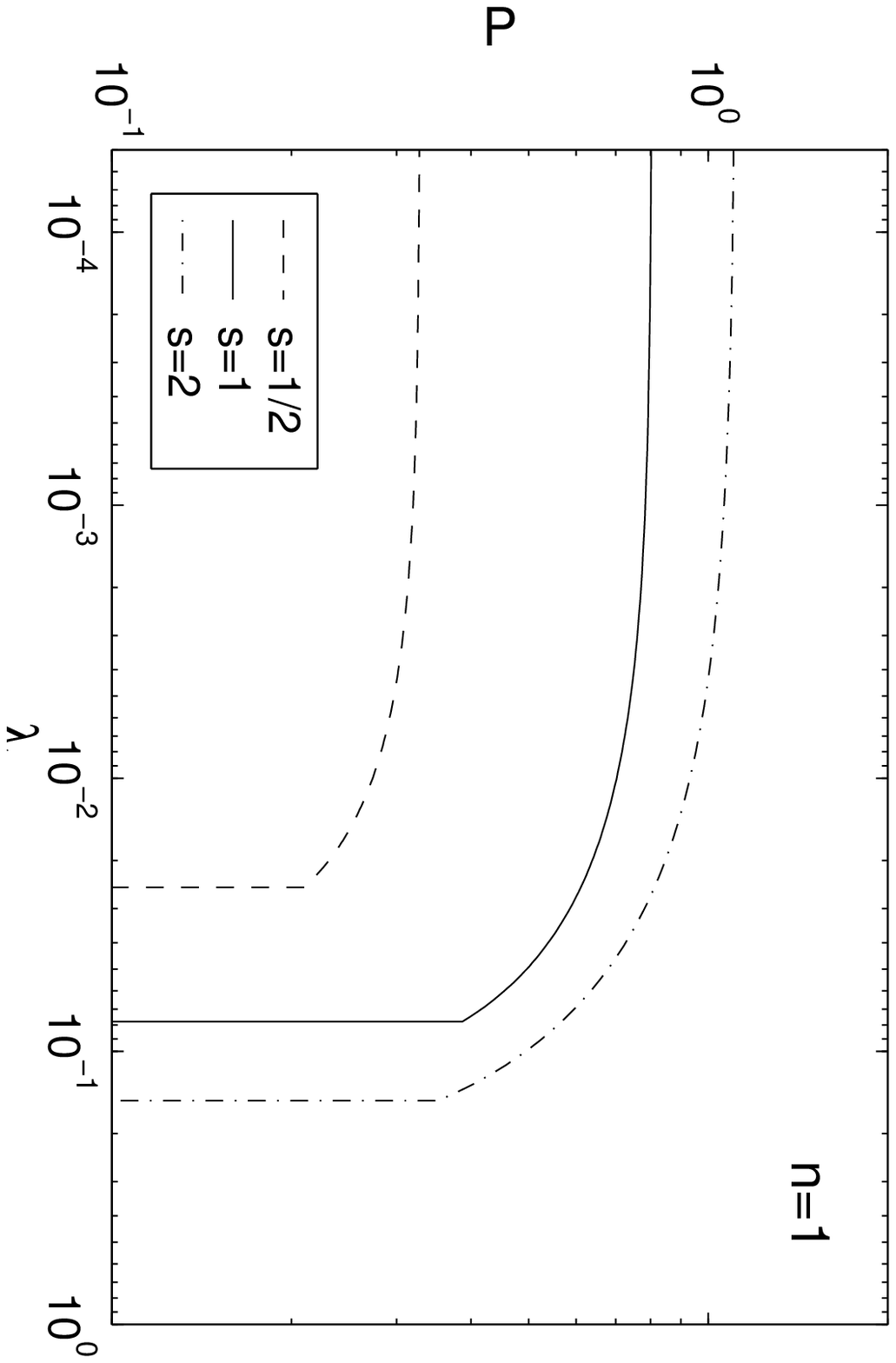,height=3.0in,width=2.5in}}
\mbox{\psfig{figure=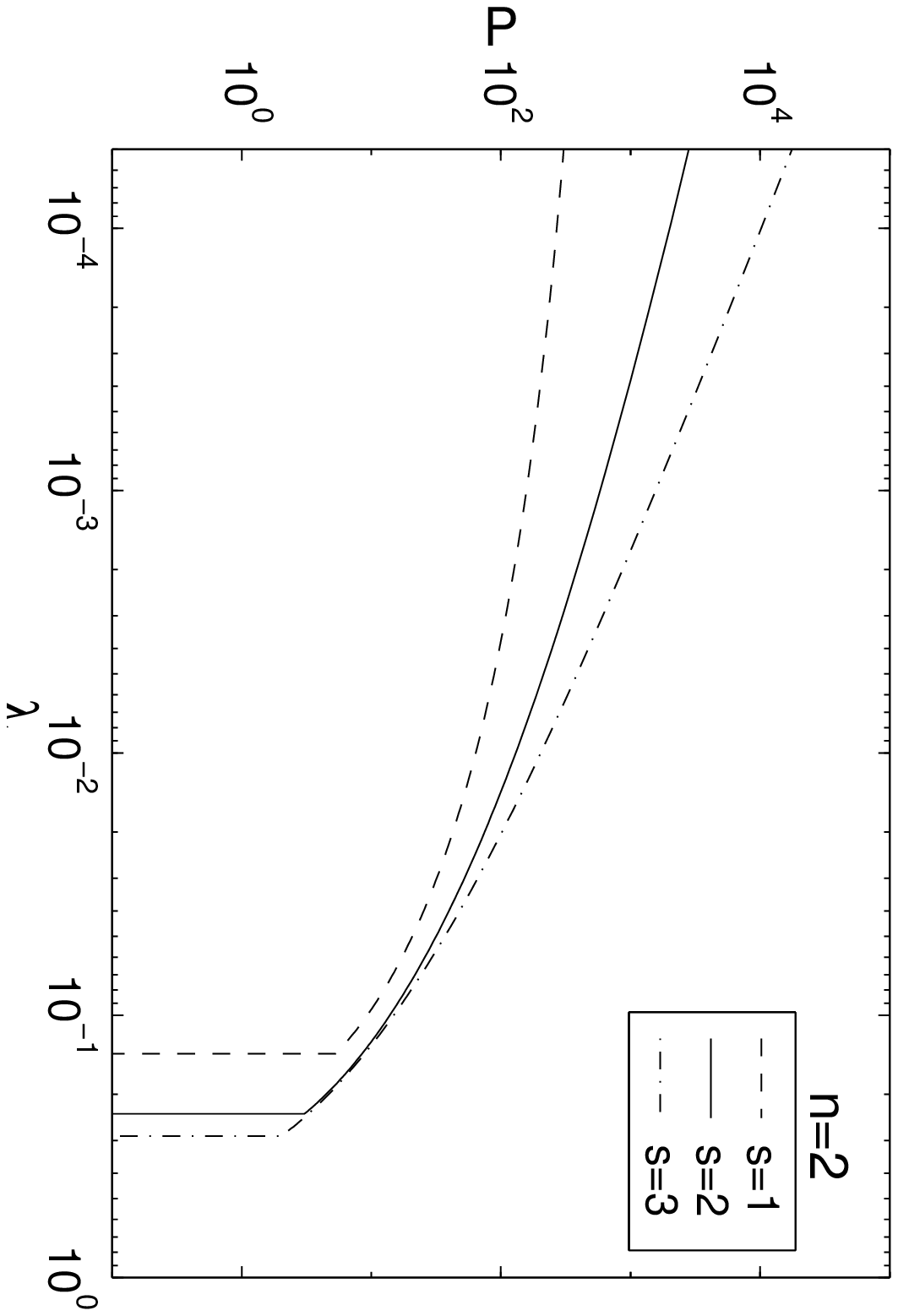,height=3.0in,width=2.5in}}
\mbox{\psfig{figure=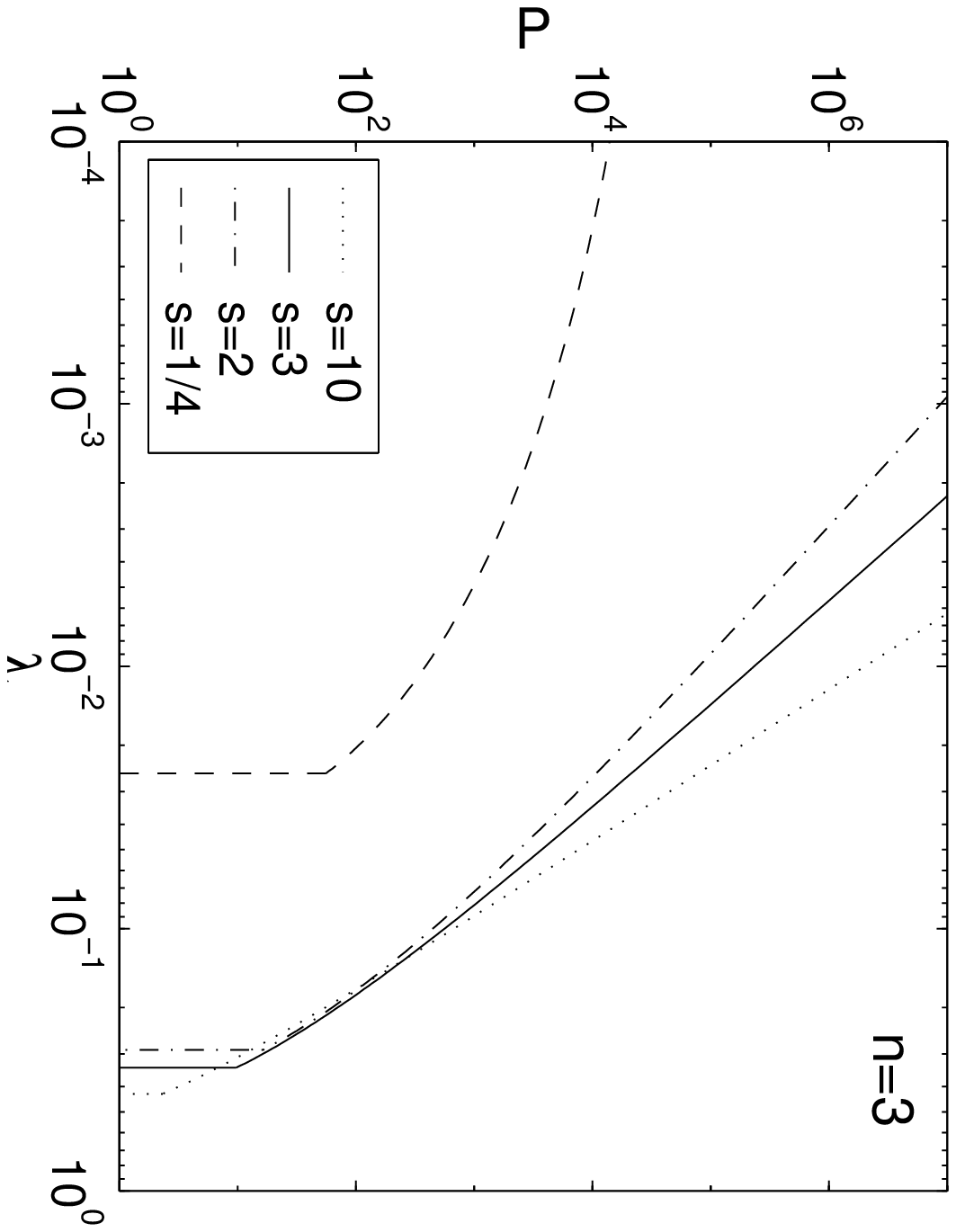,height=3.0in,width=2.5in}}
\end{sideways}
\caption{The pressure $P$. 
}
\label{fig:pres}
\end{figure}

\begin{figure}
\centering
\begin{sideways}
\mbox{\psfig{figure=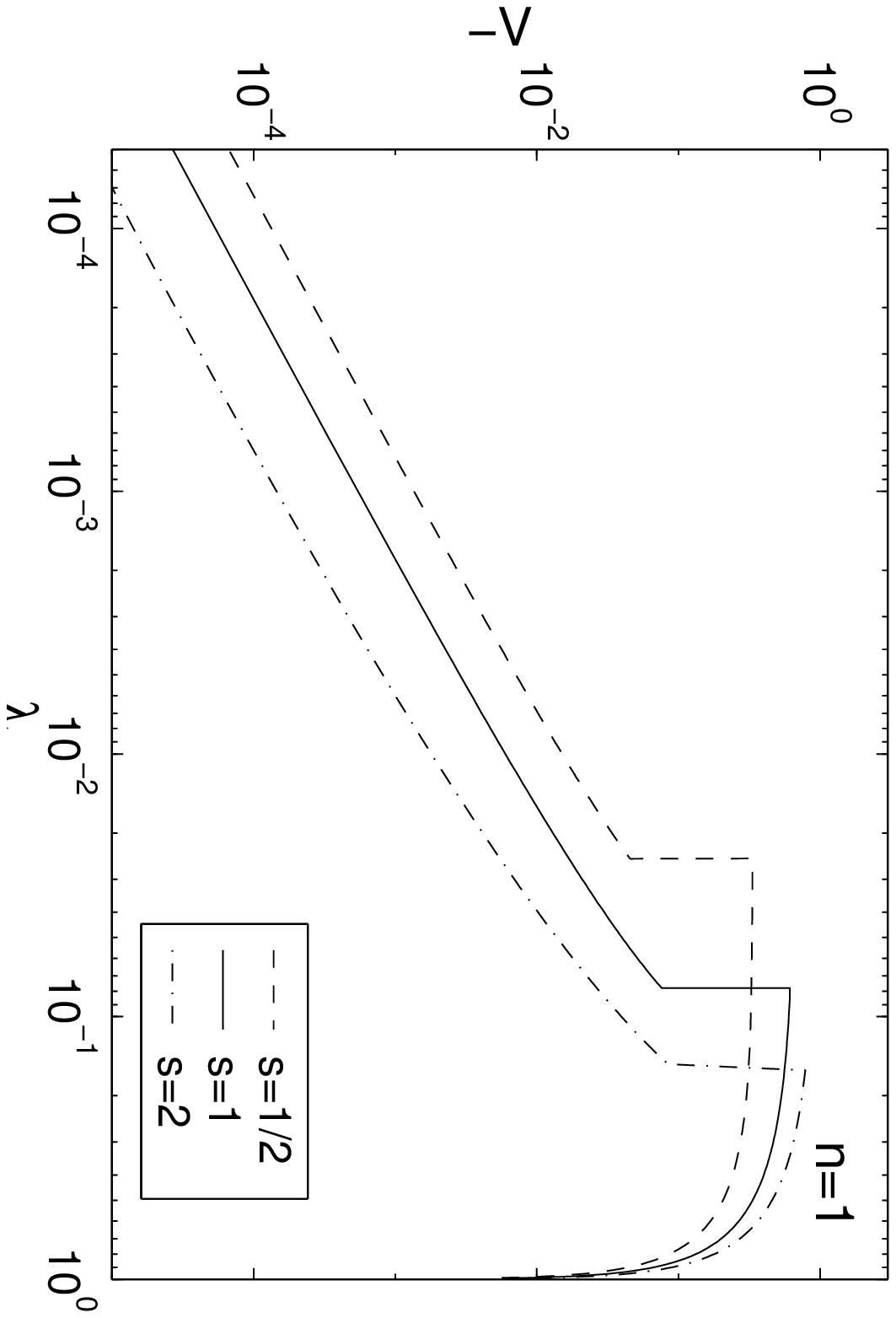,height=3.0in,width=2.5in}}
\mbox{\psfig{figure=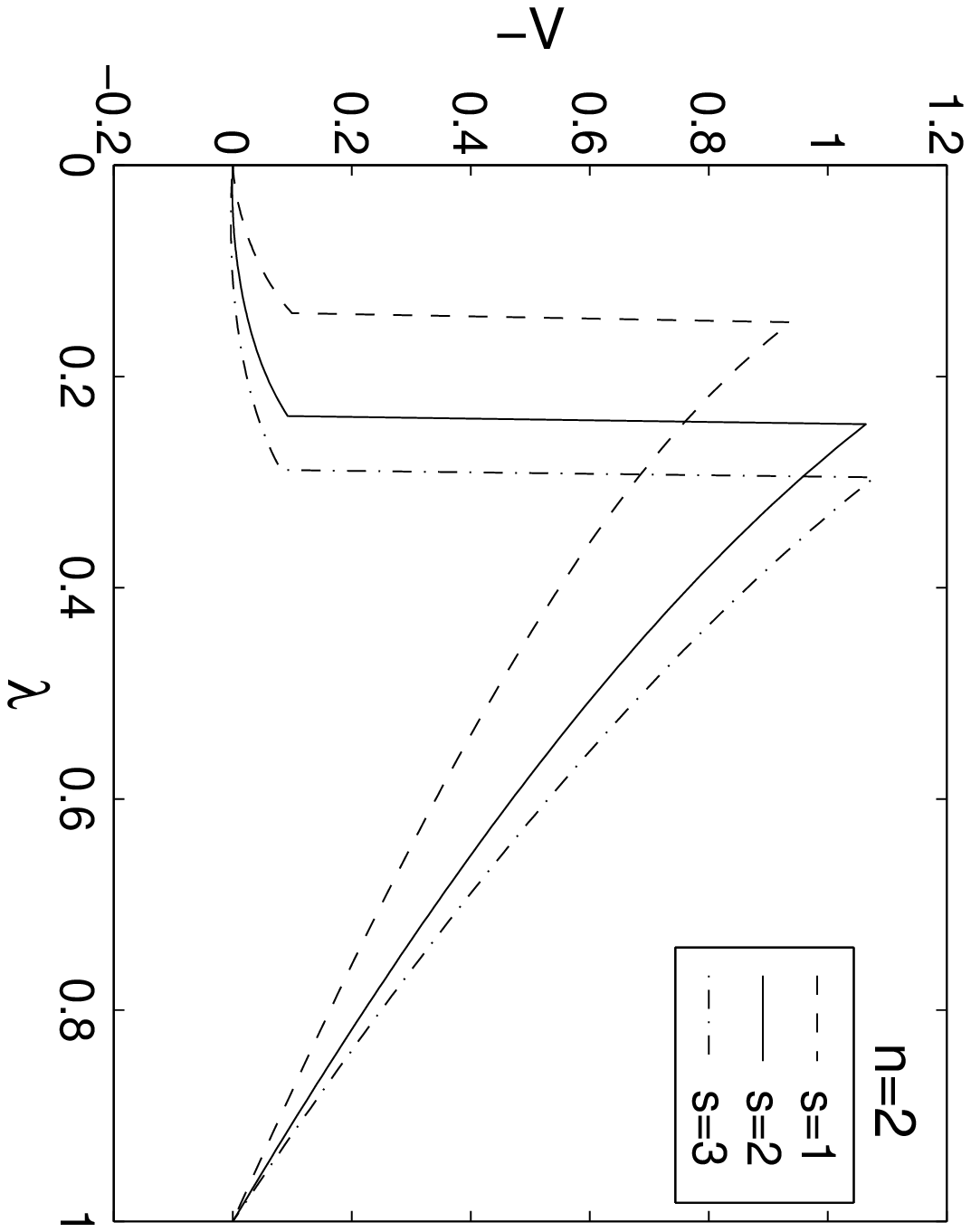,height=3.0in,width=2.5in}}
\mbox{\psfig{figure=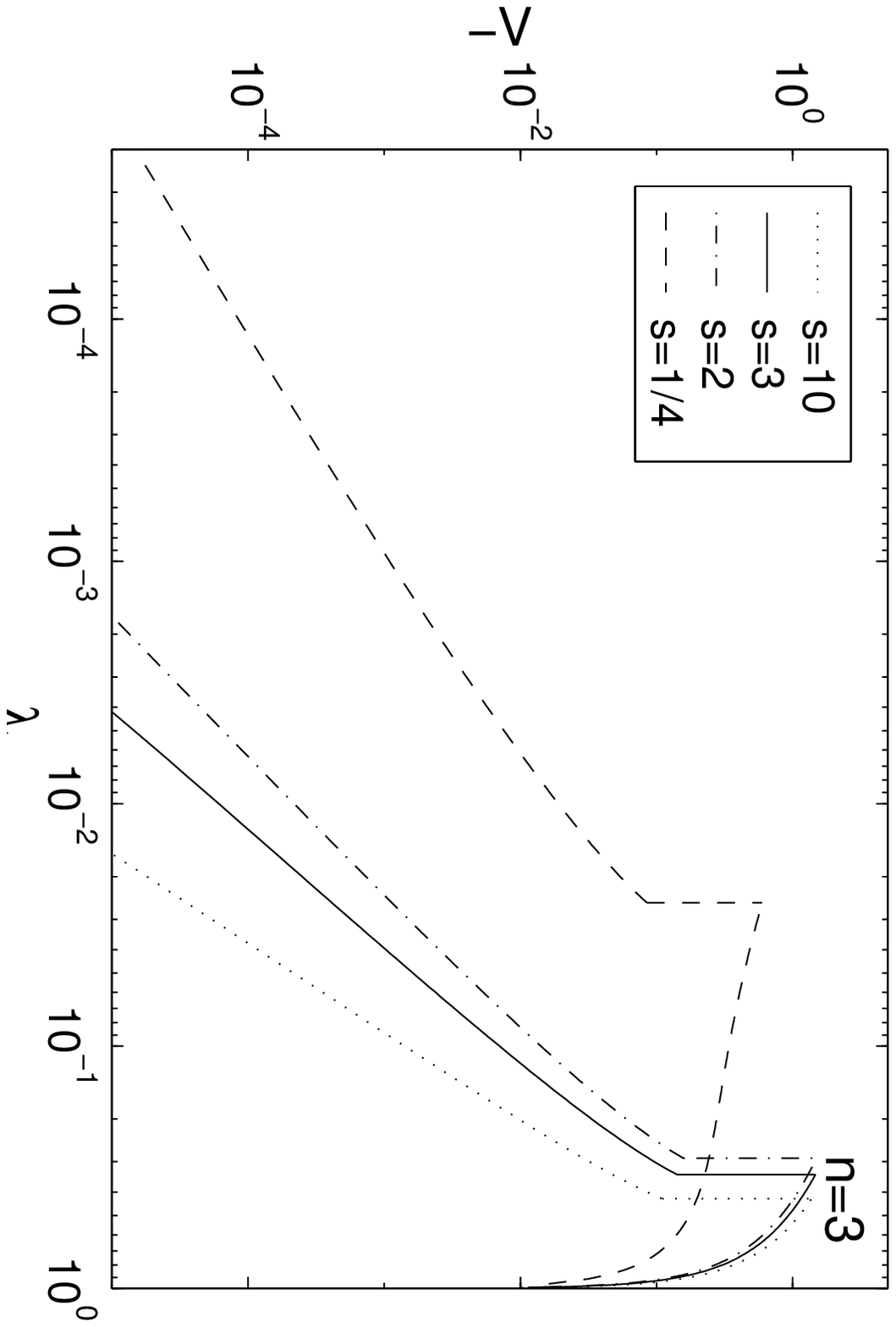,height=3.0in,width=2.5in}}
\end{sideways}
\caption{Curves of {\it minus} the velocity $V$. Linear  vertical scale
for $\nn=2$.
}
\label{fig:vel}
\end{figure}

\begin{figure}
\centering
\begin{sideways}
\mbox{\psfig{figure=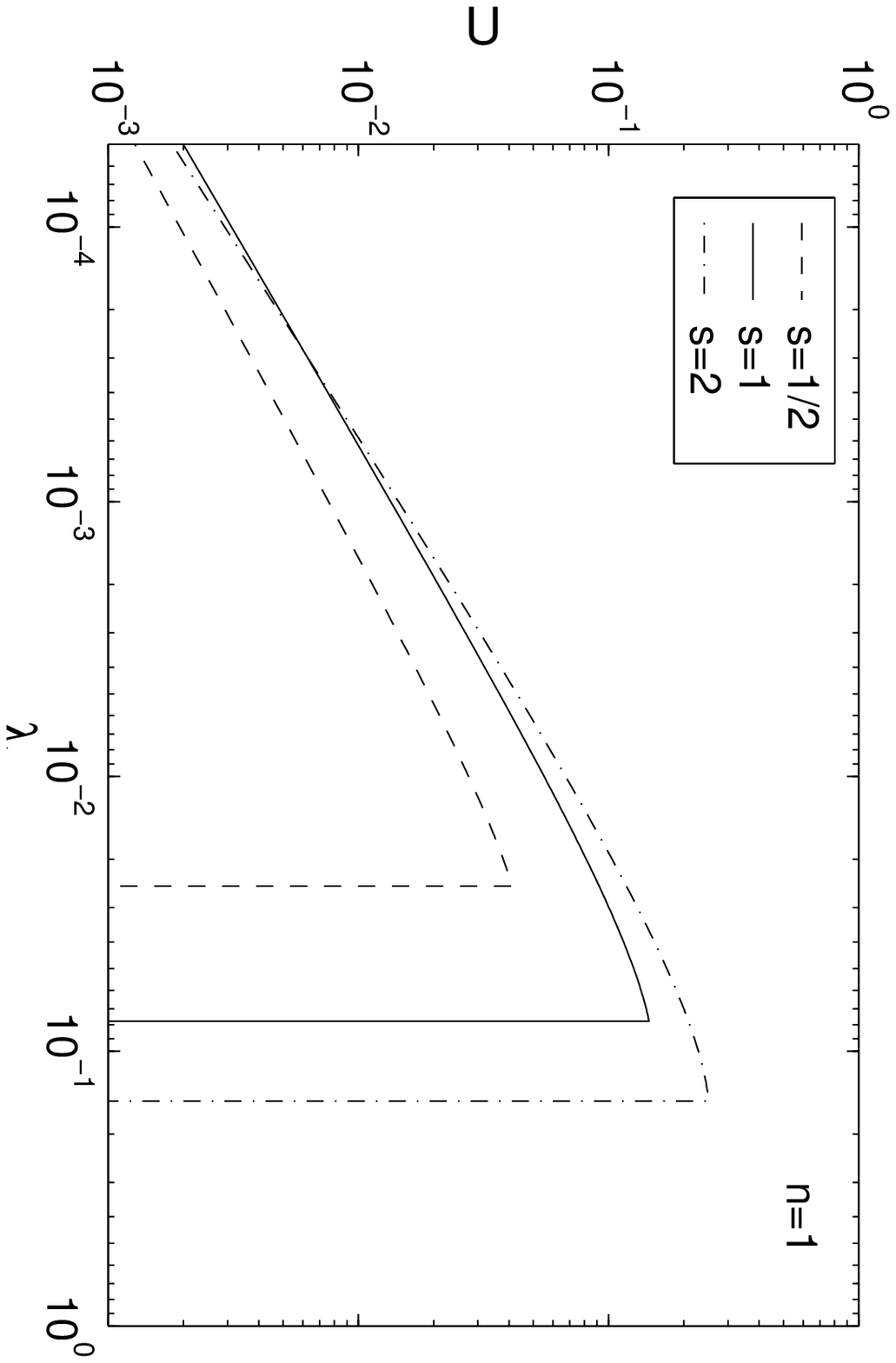,height=3.0in,width=2.5in}}
\mbox{\psfig{figure=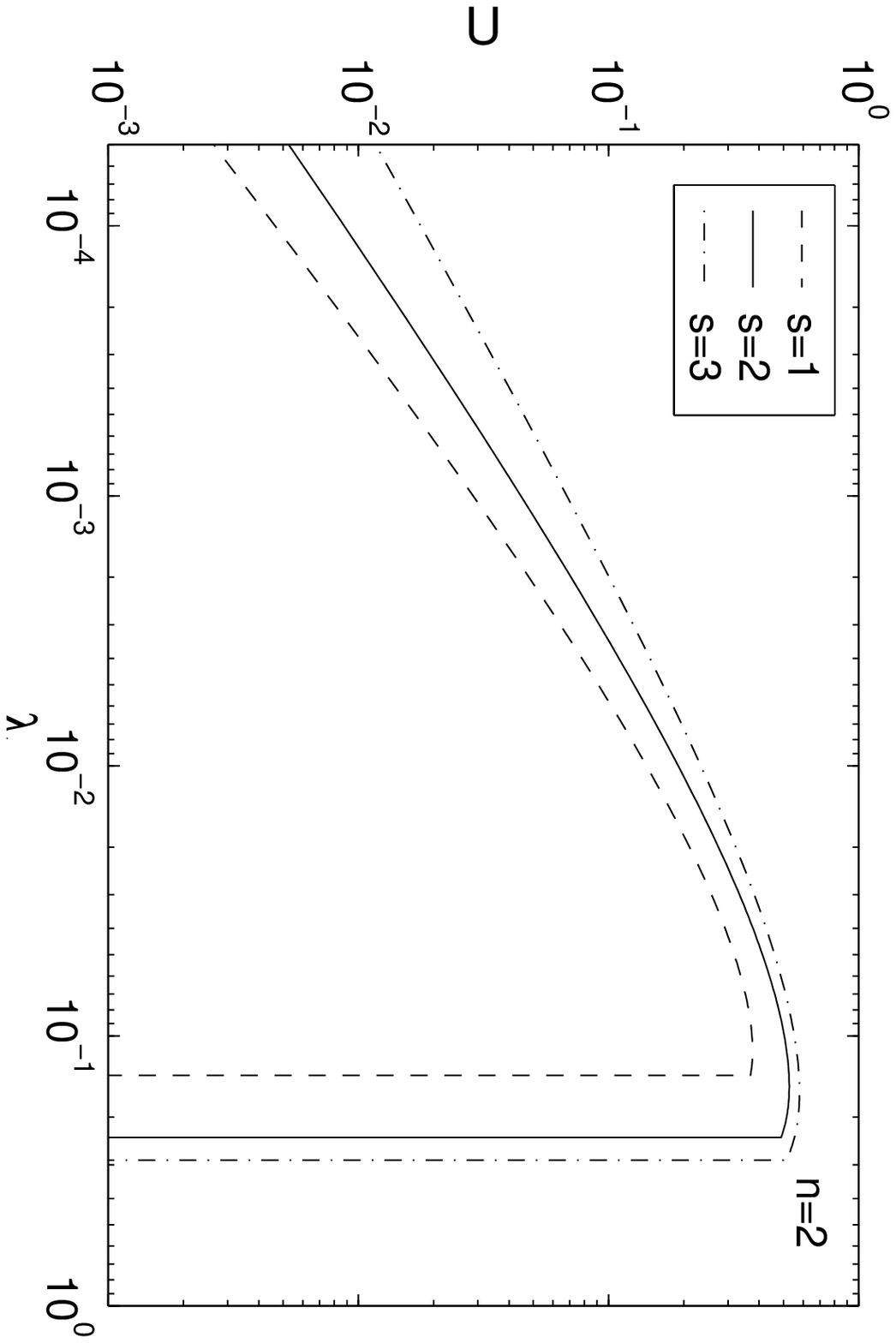,height=3.0in,width=2.5in}}
\mbox{\psfig{figure=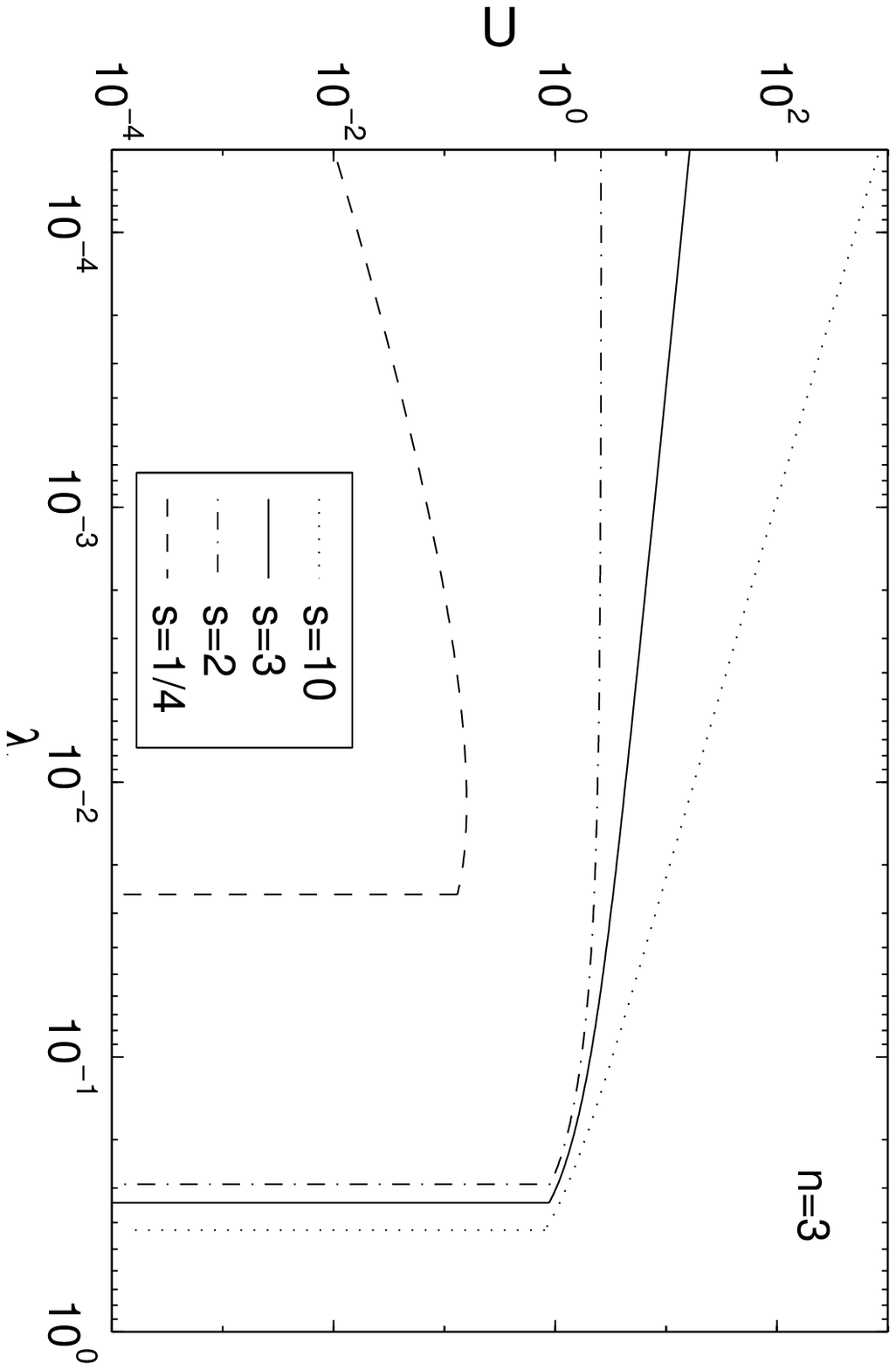,height=3.0in,width=2.5in}}
\end{sideways}
\caption{
The thermal energy $U=\frac{P}{D(\gamma-1)}$.
}
\label{fig:term}
\end{figure}

\begin{figure}
\centering
\begin{sideways}
\mbox{\psfig{figure=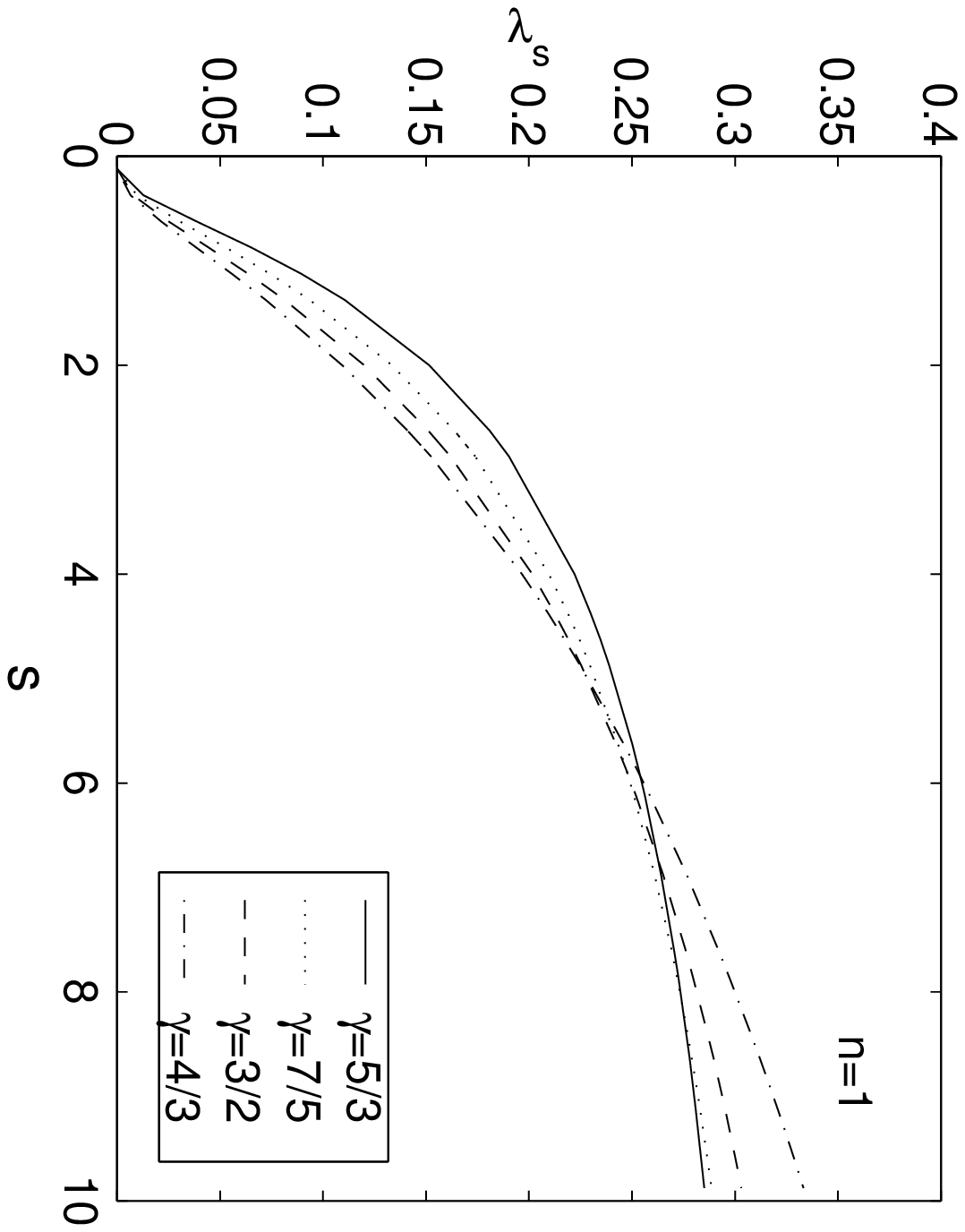,height=3.0in,width=2.5in}}
\mbox{\psfig{figure=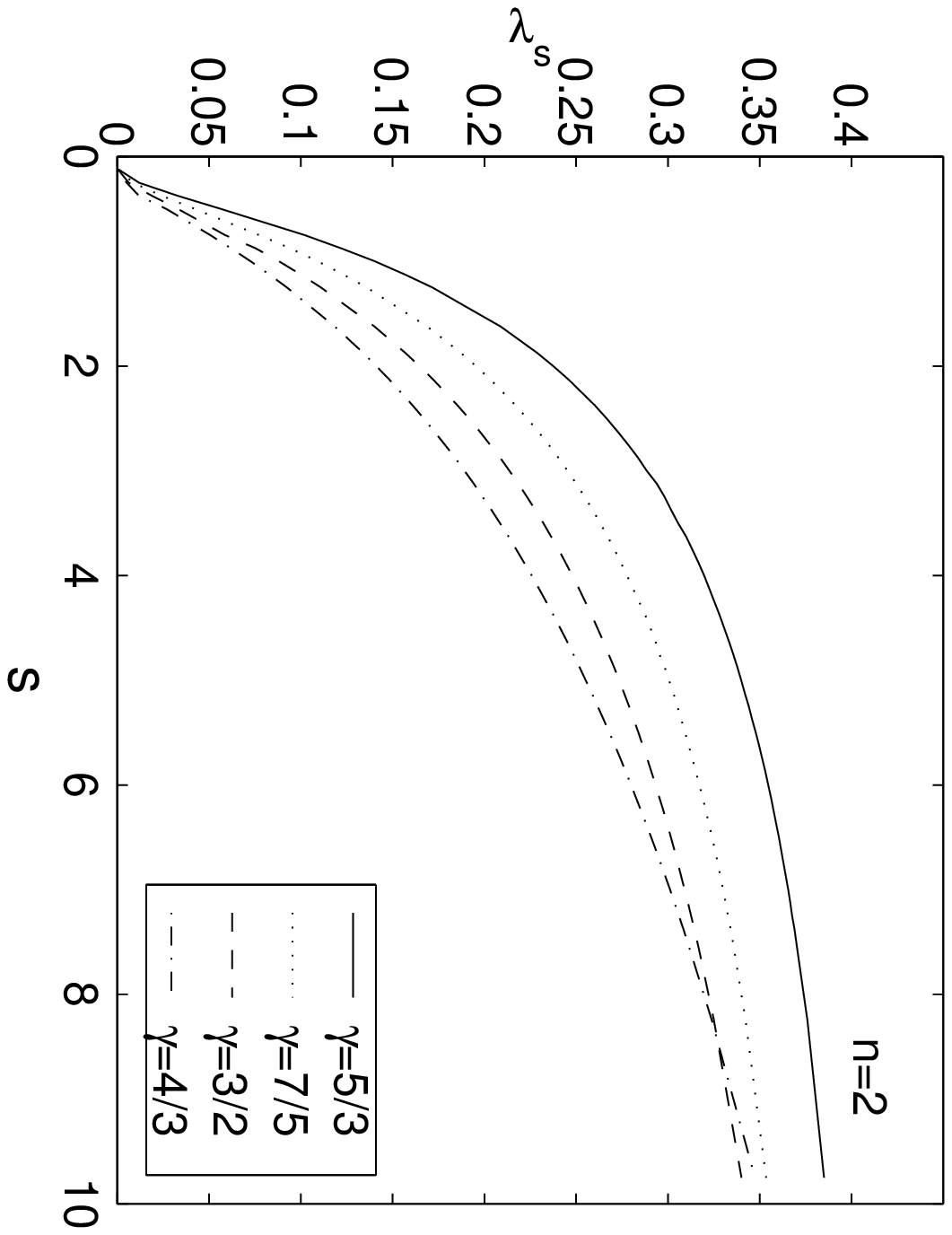,height=3.0in,width=2.5in}}
\mbox{\psfig{figure=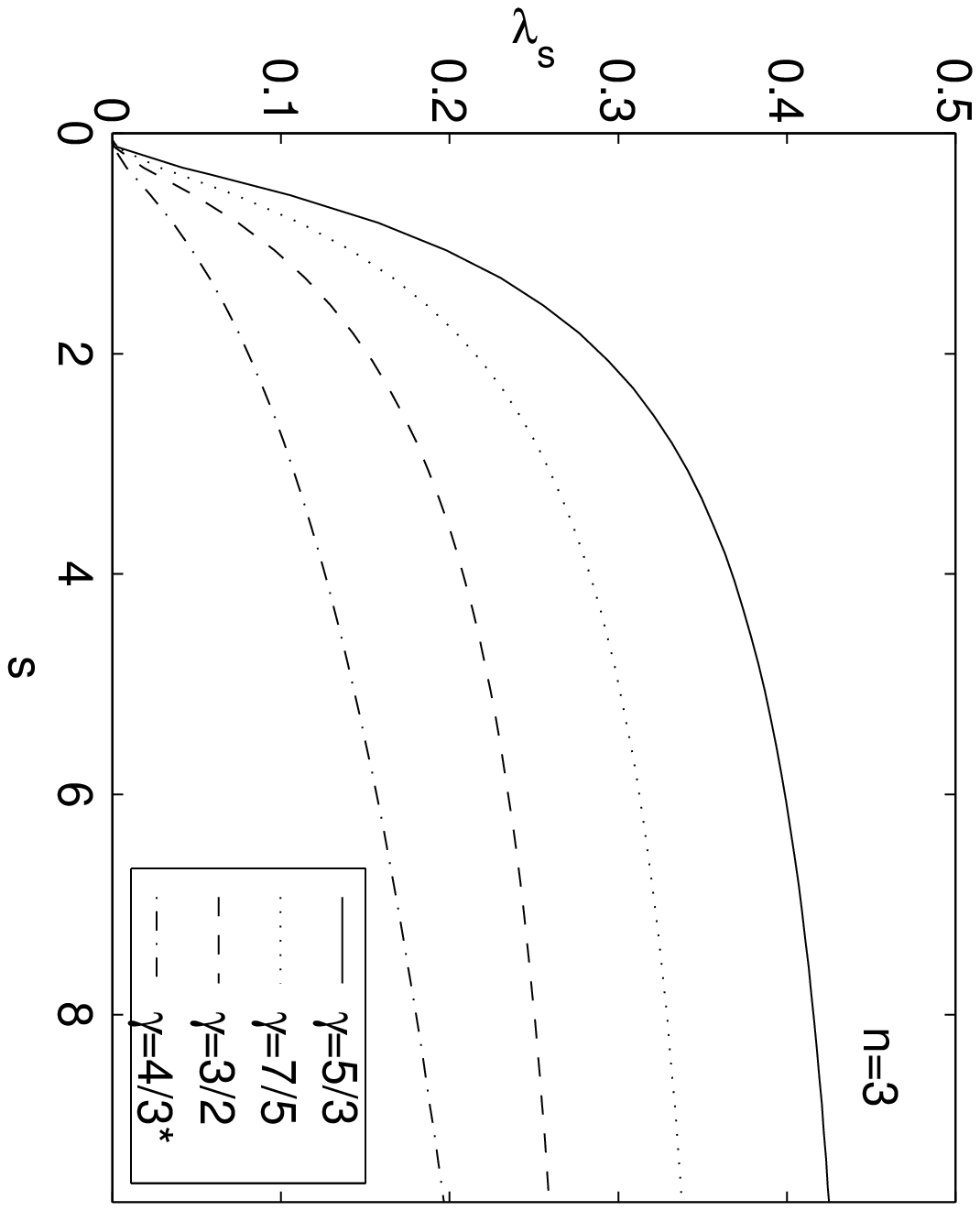,height=3.0in,width=2.5in}}
\end{sideways}
\caption{
The shock location $\lambda_s$ as a function of $\sp$.
}
\label{fig:lam}
\end{figure}

\begin{figure}
\centering
\begin{sideways}
\mbox{\psfig{figure=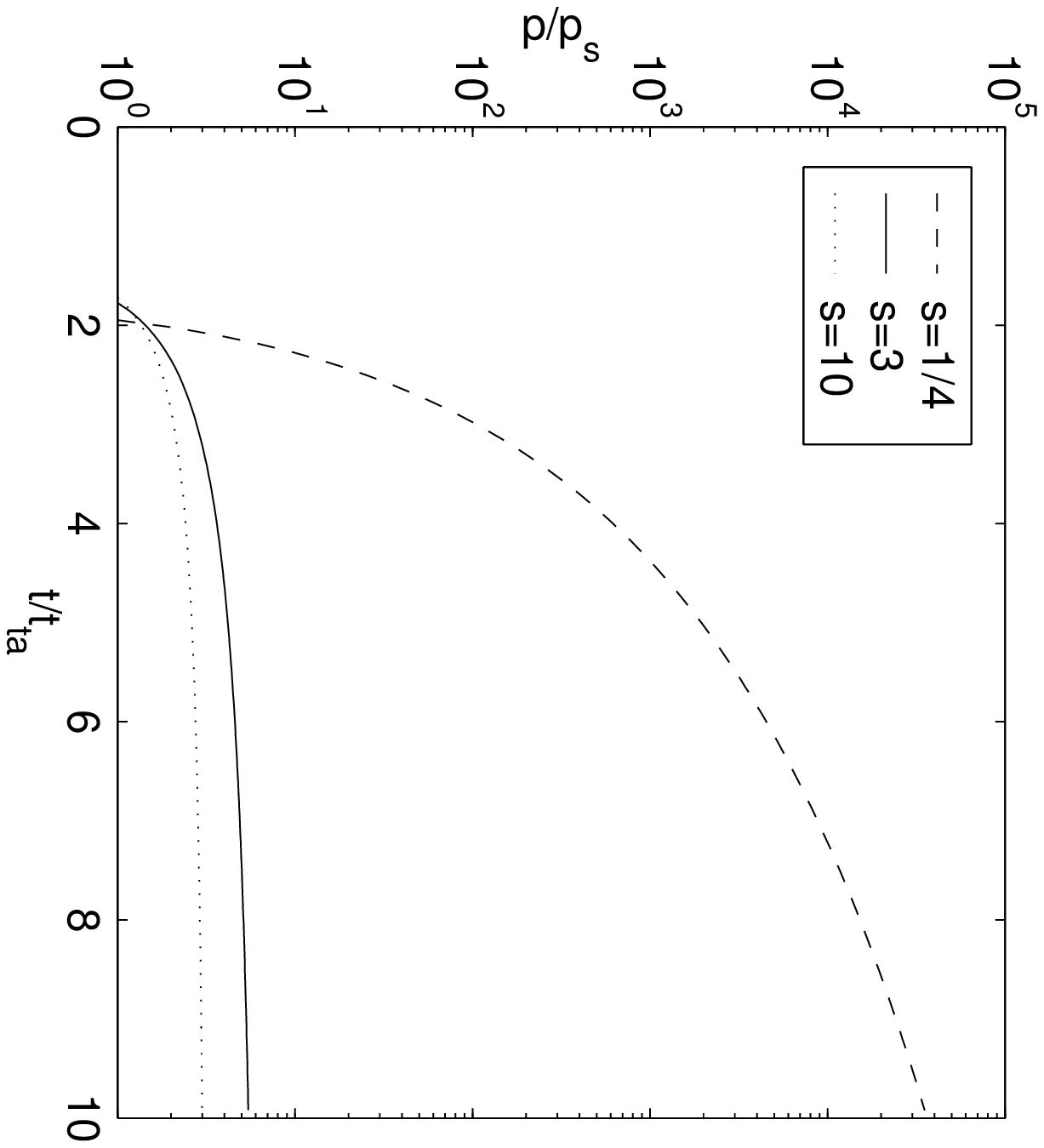,height=3.0in,width=2.5in}}
\mbox{\psfig{figure=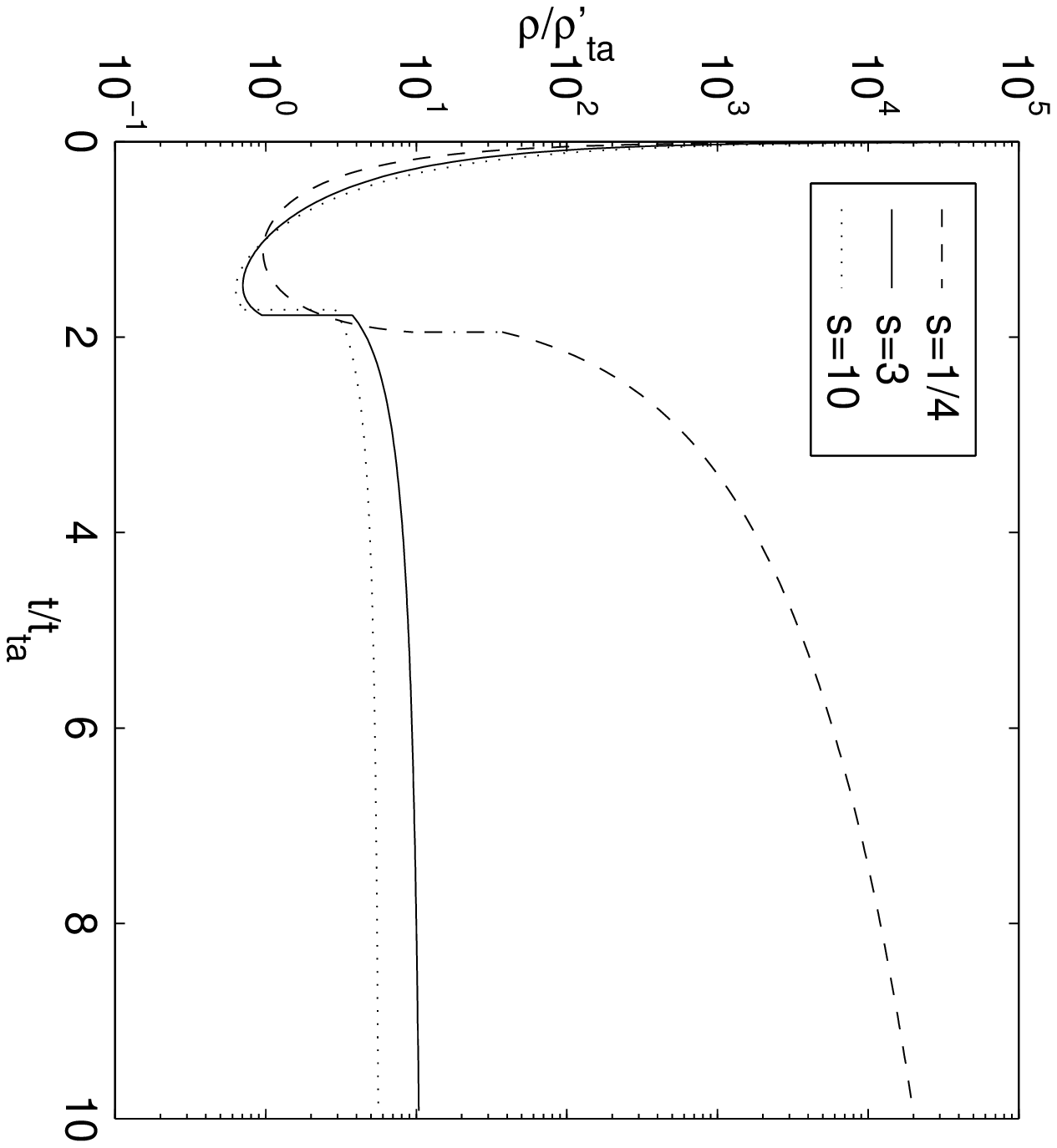,height=3.0in,width=2.5in}}
\mbox{\psfig{figure=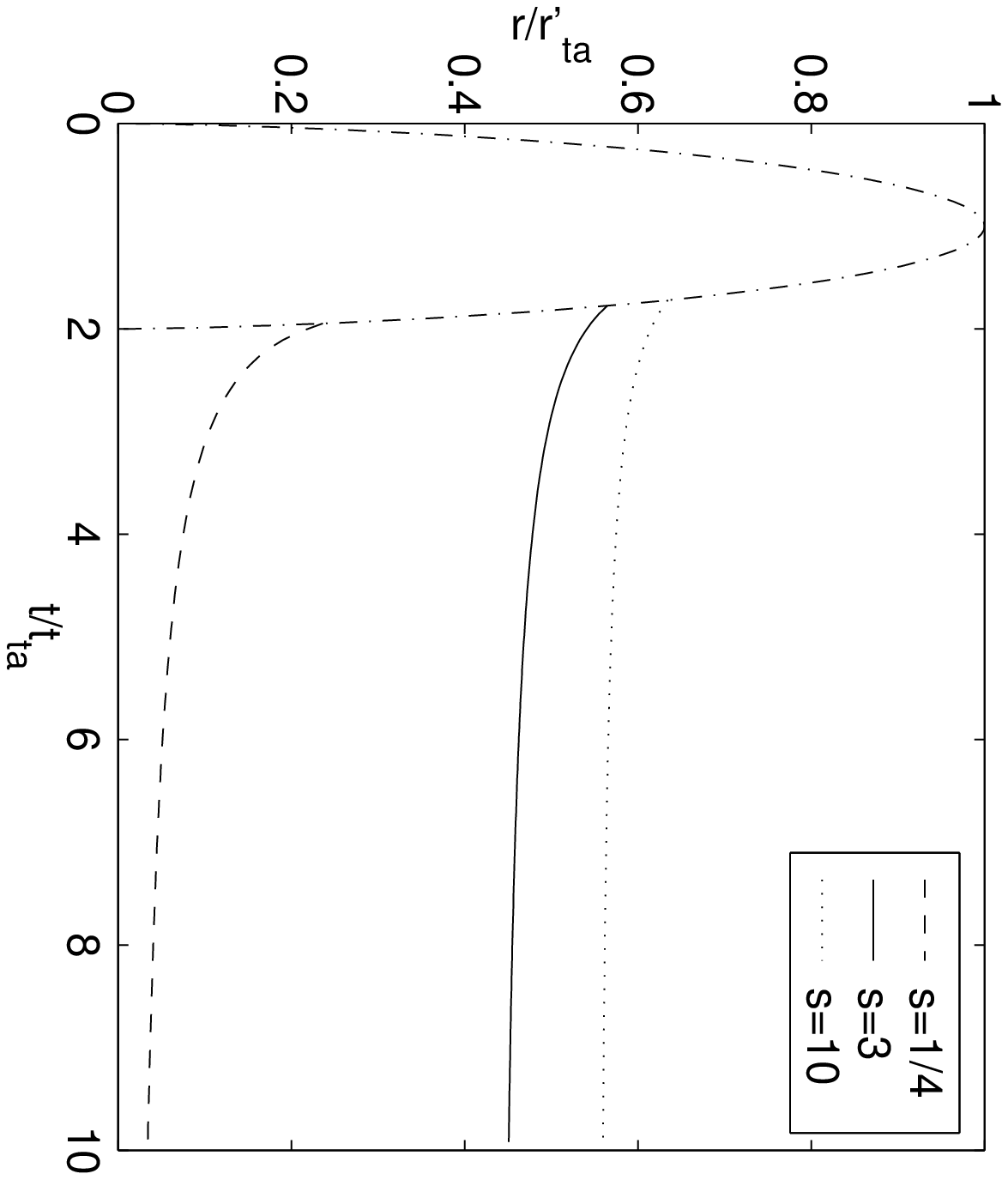,height=3.0in,width=2.5in}}
\end{sideways}
\caption{ The trajectory and fluid variables of a particle as a
function of time in spherical collapse with $\gamma=5/3$ for various
$\sp$.  The distance, $r$, and density, $ \rho$, have been scaled by
their respective values, $r'_{ta}$, and, $ \rho'_{ta}$, at the
turnaround time $t_{ta}$, while the pressure, $p$, by its value,
$p_s$, immediately after the particle crossed the shock.  }
\label{fig:traj}
\end{figure}

So far we have considered similarity solutions for collapse involving
gas only.  The collapse of scale free symmetric perturbations in a an
Einstein-De Sitter universe containing a mixture of gas and
collisionless matter is also self-similar. Similarity solutions for
mixed collapse are beyond the scope of the present paper. Here we only
obtain the asymptotic behaviour of the gas variables in a universe
dominated by collisionless matter. For a self-similar collapse to
develop, the two matter components must start with the same initial
conditions with zero initial gas pressure.  Shells of gas and
collisional matter then move together until they reach either a shock
in the gas or the region of shell crossing in the collisionless
component. The evolved density profiles of both components depend on
$r$ and $t$ through $\lambda$. Although the global gas mass fraction
is negligible, the gas can be gravitationally dominant at the center
of the collapse if it has a steeper density profile than the
collisionless matter.  For the purpose of deriving the asymptotic
exponents we proceed assuming that the gravity of the gas is
negligible everywhere and check the consistency of this assumption
according to the results.  So we replace the gas mass $M(\lambda)$ in
the non-dimensional Euler equation (\ref{a2}) by the collisionless
matter mass $\tilde M(\lambda)$.  Writing $\tilde M\sim
\lambda^{{\tilde \delta}+\nn}$ near $\lambda=0$, Fillmore \& Goldreich
(1984) find that $\tilde \delta$ is
\begin{equation}
\label{fg1}n=1: \qquad \tilde \delta = \frac{-\sp}{\sp+1}
\end{equation}
\begin{equation}
\label{fg2} n=2: \qquad \tilde \delta = -1
\end{equation}
\begin{eqnarray}
n=3: \qquad \tilde \delta &=& -2 \quad ,\quad {\rm for} \quad \sp\le 2
\nonumber \\
\label{fg3} &=& \frac{-3 \sp}{\sp+1} \quad ,\quad {\rm for} \quad \sp> 2 
\end{eqnarray}
Table 2 summarizes the values of the asymptotic constants obtained by
substituting the asymptotic behaviour of the collisionless mass in the
non-dimensional Euler equation.  The dotted and dashed curves in
figure \ref{fig:delta} represent the density exponent vs $\sp$
computed from table 1 and table 2, respectively, for $\gamma=5/3$.
The solid curve is the collisionless matter exponent $\tilde \delta$
versus $\sp$.

In spherical perturbations we see from the tables and figure
\ref{fig:delta} that the asymptotic gas density profile in a universe
with collisionless matter is steeper than that containing gas only.
For $\sp >2$ the density asymptotic exponents of the gas and
collisionless matter are identical. For $\sp <2$ the collisionless
matter density profile is steeper. This is consistent with neglecting
the gravity of the gas near the center.  As $\gamma\rightarrow 4/3$
gas density exponent approaches $-2$ for $\sp<2$ and $-3\sp/(1+\sp) $
otherwise, so the gas has the same asymptotic profile as the
collisionless matter for all $\sp$.

In cylindrical perturbations with $\sp$ large enough, the gas has a
steeper density profile than the collisionless matter.  This means
that virial motions in the collisionless matter are more effective at
balancing gravity than the gas pressure force.  So for large $\sp$ the
derivation of the asymptotic exponent in section \ref{asymptotic} is
more suitable.  A steeper gas profile also occurs in planar
perturbations with $\sp <1/(3-\gamma)$. However the asymptotic
constants obtained here and in section \ref{asymptotic} are identical
in planar geometry.

In spherical perturbations with $\sp>2$, the density $\rho(r,t)$ is
constant with time and the temperature $\sim p/\rho$ diverges like
$r^{\frac{2-\sp}{1+\sp}}$ as $r\rightarrow 0$.  To first order in the
asymptotic expansion where the gas can be in hydrostatic equilibrium
in the dominant potential well of the collisionless matter. For
$\sp<2$ the density increases with time and the temperature is
constant near the center. For comparison, in the absence of
collisionless matter, for $2>\sp >1/2$ the temperature decreases
inward but the gas is in hydrostatic equilibrium to first order in the
asymptotic expansion.

\section{discussion}
\label{discussion}

The similarity solutions are found for collapse in a flat universe
with matter density parameter $\Omega = 1$.  Because of Birchhoff's
theorem, the solutions for spherical collapse are valid in an open
universe if the current turnaround radius is well inside the spherical
region interior to which the perturbation is bound.  The statement is
incorrect for planar and cylindrical perturbations because of the
explicit appearance of cosmology dependent terms in the equations of
motion (\ref{eom1}--\ref{eom4}), like $t^{2(3-n)/3}$ in the continuity equation (\ref{eom1}).

The solutions are appropriate for the adiabatic collapse of
perturbations with deep gravitational potential so that the initial
thermal energy of the gas can be ignored. Such perturbations are
probably the seeds for massive galaxies, galaxy groups and clusters.
In the intergalactic medium (IGM) most of the gas is continuously
photo-heated and is of moderate density. There is considerable
interest in analytic modelling of the IGM in current methods for
extracting cosmological information from the Lyman forest (Croft
et. al. 1998, Nusser \& Haehnelt 1999, 2000).  So far these methods
have heavily relied on linear analysis (e.g., Bi, B\"orner, \& Chu
1992, Gnedin \& Hui 1998, Nusser 2000) and hydrodynamical simulations
(e.g., Petitjean et. al. 1995, Theuns et. al. 1999).  Analytic
treatment of the IGM beyond the linear regime is exceedingly
complicated.  Consider a situation in which photo-heating establishes
the relation $p=k \rho^\gamma$ in the IGM, where $k$ and $\gamma$
depend non-trivially on time (e.g., Theuns et. al. 1999).  The
pressure in this case introduces a length scale $k^{1/2}
G^{(1-\gamma)/2} t^{2-\gamma}$ (e.g., Sedov 1959). If we take constant
$k$ and $\gamma$, this length scale varies with time like $r_{ta}$
only in the special case of $\gamma=4/3$ and infinite $\sp$. So
physically interesting situations including initial pressure 
in which the collapse is
self-similar do not exist.

 Spherical Perturbations with $\sp>2$ when $\gamma>4/3$, and with
any $\sp>0 $ when $\gamma=4/3$ deserve special attention.  In the
corresponding solutions for shocked collapse without a black hole at
the center, the quantity $V_c^2=2Gm/r$ diverges towards the center, and
so there is a point $r=r_g$ at which $V_c^2=c^2$, where $c$ is the
speed of light. According to general relativity this implies the presence
of a black hole at the center, invalidating the assumption of no
central mass, made in deriving the solutions.  In particular, the
condition $\upsilon(r=0)=0$, necessary for shocked collapse, is
incompatible with the presence of a central black hole. 
Near $r_g$, however,   radiation pressure
and angular momentum can prevent
 the formation of a black hole.  
Should this occur, 
we expect our solution for shocked accretion to be valid away
from the central region. 

The evolved gas variables in the symmetric self-similar collapse
contain full information on the initial perturbation.  So the system
retains memory of the initial conditions, even in the highly nonlinear
regime.  On the other hand, a collapsing system of collisionless
matter can develop density profiles which do not depend on the initial
shape of the perturbation. For example, according to the solutions of
Fillmore \& Goldreich (1984), a spherical density perturbation
develops into $r^{-2}$ for $\sp<2$, and a cylindrical perturbation
into $r^{-1}$ for all $\sp$.  Haloes identified in cosmological
simulations of collisionless particles with generic initial
conditions, also tend to have density profiles independent of the
spectrum of the initial fluctuations (Navarro, Frenk \& White 1997).

Our results are relevant for describing the gas distribution in
various physical systems such as the cores of clusters or pancake-like
superclusters.  Over a limited range of scales, the index $\sp$ can be
related to the index, $l$, of the three dimensional power spectrum,
$p(k)\sim k^l$, of the linear density fluctuations.  If the initial
density field is gaussian with a scale free power spectrum then the
properties of the nonlinear field depend only on one scale. This is
the nonlinear scale, $R_{nl}$, defined as the scale on which the rms
value of density fluctuations is unity.  This scale grows with time
like\footnote{$R_{nl}$ does not involve the dimension $\nn$ because
$l$ refers to the three dimensional $p(k)$ so the rms value on a scale
$R$ is $R^{(l+3)/2}$ independent of $\nn$. } $R_{nl}\sim
t^{\frac{2(l+5)}{3(l+3)}}$.  By matching the time dependence of
$R_{nl}$ and $r_{ta}\sim t^{\frac{2(\sp+1)}{3\sp}}$ we identify
$\sp=(l+3)/2$.  So the collapse of gas into galaxies and clusters can
be, respectively, modeled by our solutions for $ \sp\!\sim\! 0.4$ and
$0.65$, where we have taken $l\!\sim\!  -2.2$ and $-1.7$ assuming a
Cold Dark Matter power spectrum.  Taking $l\!\sim\! -1$ for collapse
on a pancake-like large scale superclusters gives $\sp\sim 1$.
Another way to relate $\sp$ and $l$ is to identify symmetric
perturbations with local maxima in the linear density field (Hoffman
\& Shaham 1985).  The shape of high density peaks in a gaussian field
varies with $r$ like the two-point correlation function, $\sim
r^{-(l+3)}$ (e.g., Bardeen et. al. 1986).  So, at least in the limit
of high peaks, $\sp=l+3$. On galaxy and cluster scales the relation
$\sp=l+3$, respectively, gives $\sp=0.8$ and $1.3$ in contrast to
$\sp=0.4 $ and $0.65$ obtained from $\sp=(l+3)/2$.  Note however that
gas in galaxies tends to settle into disks and so our solutions are
less relevant than they are for cluster size objects.

In spherical geometry the asymptotic behaviour shows that the gas
cannot be pressure supported if $\sp<1/2$, and $\sp<2$ for collapse
with, and without collisionless matter, respectively.  Estimates of
the masses of rich galaxy clusters from X-ray observations of the
intracluster gas rely on hydrostatic equilibrium (e.g., Fabian 1994).
If on cluster scales $\sp\sim 0.7$--$\sim 1.3$, then the asymptotic
behaviour implies that the cluster gas may not be in hydrostatic
equilibrium.  How large is the error introduced in the mass estimates
by assuming hydrostatic equilibrium?  The following argument shows
that this error is negligible.  Hydrostatic equilibrium calculations
neglect the term $ G^{-1}r^2\dd \upsilon /\dd t$ in the mass estimate.
Using the asymptotic expansion one finds that neglecting this term
amounts to a relative mass error of $\sim (2/\pi^2)
(t/t_{ta})^{3V_0-2}$ where $t_{ta}$ is the turnaround time of the
shell present at $r$ at the current time $t$.  Shells in the inner
regions have passed their maximum expansion a few dynamical times
ago. Therefore $t\gg t_{ta}$ and since $V_0<0$ we conclude that the
error is negligible.
 
The solutions are  related to modelling the structure of
haloes made of self interacting dark matter (SIDM) (Spergel \&
Steinhardt 1999) with large interaction cross section.  On scales of
massive galaxies and clusters, our results predict final density
profile $\rho\sim r^{-1.2}$ to $\sim r^{-1.7}$.  These profiles are
consistent with the results obtained by Moore et. al.  in their
simulations of SIDM with large cross section.

\section{acknowledgement}
We thank the referee for useful comments.  This research was supported
by The Israel Science Foundation founded by The Israel Academy of
Sciences and Humanities, and by the Fund for the Promotion of Research
at The Technion.

\eject
\begin{table}
\caption{ Asymptotic constants, $V_0$, $\delta$, and $\eta$ for
collapse of collisional gas only. Spherical perturbations with
$\gamma=4/3$ allow a range of $\lambda_s$. Listed are the values
corresponding to the maximal $\lambda_s$. }
\begin{tabular}{l|c|c|c|}
&& $\nn=3,\; \gamma>4/3 $& \\  \hline
&$V_0$ & $\eta $& $\delta$\\  \hline
$\sp\le 1/2 $&$\frac{4(2s-1)}{9s\gamma} $&$ 0 $& $\frac{-3(2+s(3\gamma-4))}{2+3\gamma+s(3\gamma-4)}$ \\ \hline 
$\sp>  1/2 $&$0 $&$2\frac{1-2\sp}{\sp+1}$& $-\frac{3\sp}{\sp+1}$\\ 
\hline\hline 
&& & \\  
&& $\nn=3,\; \gamma=4/3 $& \\  \hline
&$\frac{2(s-5)}{3s} $&-3&-2.5 \\ \hline\hline 
&& & \\ 
 
&& $\nn=2,\; \gamma\ge 4/3 $& \\ \hline 
$s\leq
\frac{2(\gamma-1)}{3\gamma-4} $&$ \frac{s(4-\gamma)-2}{3s\gamma} $&$ 0
$&$\frac{-2(2+s(3\gamma-4))}{2+2\gamma+s(3\gamma-4)} $ \\ \hline
$s>\frac{2(\gamma-1)}{3\gamma-4} $&$ \frac{2-\gamma}{3(\gamma-1)}
$&$\frac{4(1-\gamma)+2s(3\gamma-4)}{2(1-\gamma)+s(4-3\gamma)} $&$
\frac{2s(4-3\gamma)}{2(\gamma-1)+s(3\gamma-4)}$ \\ \hline \hline

&& & \\ 
&& $\nn=1,\; \gamma\ge 4/3 $& \\  \hline
 &$ \frac{-4(1+s(\gamma-2))}{3s\gamma} $&$ 0 $& $\frac{s(4-3\gamma)-2}{2+\gamma+s(3\gamma-4)}$ \\ \hline 
\end{tabular}
\label{tab:3}
\end{table}
\begin{table}
\caption{
The gas asymptotic constants, $V_0, \delta,\eta$ for mixed
collapse with dominant collisionless matter.}
\begin{tabular}{l|c|c|c|}
&& $\nn=3,\;\gamma\ge 4/3  $& \\  \hline
&$V_0$ & $\eta $& $\delta$\\  \hline
$\sp\le 2 $&$\frac{2(s-2)}{9s(\gamma-1)} $&$\frac{2+s(3\gamma-4)}{(1+4s)/3-\gamma(1+s)} $&$\frac{2+s(3\gamma-4)}{(1+4s)/3-\gamma(1+s)}  $ \\ \hline 
$\sp>2 $&$0 $&$2\frac{1-2\sp}{\sp+1}$& $-\frac{3\sp}{\sp+1}$\\ 
\hline\hline 

&& & \\  
&& $\nn=2,\; \gamma\ge 4/3 $& \\  \hline
$s\leq \frac{2(\gamma-1)}{3\gamma-4} $&$ \frac{s(4-\gamma)-2}{3s\gamma} $&$ 0 $&$\frac{-2(2+s(3\gamma-4)}{2+2\gamma+s(3\gamma-4)} $ \\ \hline 
$s>\frac{2(\gamma-1)}{3\gamma-4} $&$ \frac{2(1+s(\gamma-3))}{3s(1-2\gamma)} $&$\frac{2(\gamma-1)-s(3\gamma-4)}{2\gamma+s(3\gamma-4)} $&$ \frac{-2+2s(4-3\gamma)}{2\gamma+s(3\gamma-4)}$ \\ \hline \hline

&& & \\ 
&& $\nn=1,\; \gamma\ge 4/3 $& \\  \hline
 &$ \frac{-4(1+s(\gamma-2))}{3s\gamma} $&$ 0 $& $\frac{s(4-3\gamma)-2}{2+\gamma+s(3\gamma-4)}$ \\ \hline

\end{tabular}
\label{tab:4}
\end{table}

\begin{thebibliography}{}
\bibitem{} Bardeen J.M., Bond J.R., Kaiser N., Szalay A.S., 1986, ApJ, 304, 15
\bibitem{} Benson, A.,J., Cole, S., Frenk, C.S., Baugh, C.M.,
Lacey, C.G., 2000, MNRAS, 311, 793
\bibitem{} Bertschinger E., 1985, ApJS, 58, 39
\bibitem{} Bertschinger E., 1983, ApJ, 268, 17
\bibitem{} Bi H.G., B\"orner G., Chu Y., 1992, A\&A, 266, 1    
\bibitem{}  Cole, S., Aragon-Salamanca, A.,  Frenk, C.S., Zepf, S.E., 1994,
MNRAS, 271, 781
\bibitem{} Croft R.A.C., Weinberg D.H., Katz N., Hernquist L., 1998, ApJ, 
495, 44 
\bibitem{} Dekel A., Rees M.J., 1987, Nature, 326, 455
\bibitem{} Fabian A.C., 1994, Annu. Rev. Astron. Astrophys., 32, 277
\bibitem{} Fillmore J.A.,  Goldreich P., 1984, 281, 1
\bibitem{} Flores R.A., Primack J.R., ApJL, 427, 1
\bibitem{} Forcada-Miro M.I., White S.D.M., 1997, astro-ph/9712204 
\bibitem{} Gnedin N., Hui L., 1998, MNRAS, 296, 44
\bibitem{} Gunn J.E., 1977, ApJ, 218, 592
\bibitem{} Hoffman Y., Shaham J., 1985, ApJ, 297, 16
\bibitem{} Kauffmann, G., Nusser, A., Steinmetz, M., 1997, MNRAS,286, 795
\bibitem{} Kauffmann, G., White, S.D.M., Guiderdoni, B., 1993, MNRAS, 264, 201
\bibitem{} Moore B., Gelato S., Jenkins A., Pearce F.R., 
Quilis V., 2000, astro-ph/0002308 
\bibitem{} Navarro J., Frenk C.S., White S.D.M., 1997, ApJ, 490, 493
\bibitem{} Nusser A., Haehnelt M., 1999, MNRAS, 303, 179
\bibitem{} Nusser A., Haehnelt M., 2000, MNRAS, 312, 364
\bibitem{} Nusser A., 2000, MNRAS, accepted
\bibitem{} Peebles P.J.E., 1980, {\it ``The Large Scale Structure in 
The Universe''}, Princeton University Press, Princeton.
\bibitem{} Petitjean P., M\"ucket  J.P., Kates R.E., 1995, A\&A, 295, L9
\bibitem{} Sedov L., 1959, {\it Similarity and Dimensional Methods}, London:
Cleaver-Hume press. 
\bibitem{} Somerville, R.S., Primack, J.R., 1999, MNRAS, 310, 1087
\bibitem{} Spergel D.N., Steinhardt P/J., 1999, astro-ph/9909386
\bibitem {}Theuns T., Leonard A., Efstathiou G.,
           Pearce F.R., Thomas P.A., 1998, MNRAS, 301, 478   
\bibitem{} Zel'dovich Ya.B., 1970, Astron. \& Astrophys., 5, 84
\end{thebibliography}
\end{document}